\documentclass[a4paper,left]{article}
\pdfoutput=1 

\usepackage{jcappub} 

\usepackage[T1]{fontenc} 
\usepackage[usenames,dvipsnames]{xcolor} 
\usepackage{amsmath,amssymb} 
\usepackage{epstopdf} 

\usepackage{times}
\usepackage{bm} 

\newcommand{\mr}[1]{\mathrm{#1}}

\title{\boldmath All-sky angular power spectra from cleaned WISE$\times$SuperCOSMOS galaxy number counts}

\author[a,1]{H. S. Xavier\note{Corresponding author.}}
\author[a]{, M. V. Costa-Duarte} 
\author[b,c]{, A. Balaguera-Antol\'{i}nez} 
\author[d,e]{and M. Bilicki}
\affiliation[a]{Instituto de Astronomia, Geof\'{i}sica e Ci\^{e}ncias Atmosf\'{e}ricas da Universidade de S\~{a}o Paulo,\\
Rua do Mat\~{a}o, 1226, Cidade Universit\'{a}ria, S\~{a}o Paulo, SP, Brazil}
\affiliation[b]{Instituto de Astrof\'{i}sica de Canarias, s/n, E-38205, La Laguna, Tenerife, Spain}
\affiliation[c]{Departamento de Astrof\'{i}sica, Universidad de La Laguna, E-38206, La Laguna, Tenerife, Spain}
\affiliation[d]{Leiden Observatory, Leiden University, Niels Bohrweg 2, NL-2333 CA Leiden, the Netherlands}
\affiliation[e]{Center for Theoretical Physics, Polish Academy of Sciences, al. Lotnik\'{o}w 32/46, 02-668, Warsaw, Poland}

\emailAdd{hsxavier@if.usp.br}
\emailAdd{mvcduarte@astro.iag.usp.br}
\emailAdd{balaguera@iac.es}
\emailAdd{bilicki@strw.leidenuniv.nl}

\abstract{Aiming to extract cosmological information from linear scales of the WISE$\times$SuperCOSMOS photometric redshift catalog, 
  we perform a characterization of the systematic effects associated with stellar content, evidencing the presence of contamination 
  and obscuration. We create an integrated model for these effects (which together we call ``usurper contamination''), 
  devise a method to remove both of them simultaneously and show its functionality by applying it to a 
  set of mock catalogs. When administered 
  to WISE$\times$SuperCOSMOS data, our method shows to improve the measurements of angular power spectra 
  on scales $\ell\lesssim15$ and the extraction of cosmological parameters therefrom, even though 
  a significant excess of power remains at these scales. When ignoring scales $\ell<15$, we still find 
  strong indications of systematics, albeit these can be localized in the southern equatorial hemisphere. 
  An independent analysis of the northern hemisphere at $\ell\geq 15$ agrees with the $\Lambda$CDM model with 
  parameters from the Planck satellite and gives $\Omega_{c}=0.254\pm0.020$ and 
  $\Omega_{\mr{b}}<0.065$ at 95\% confidence limit when combined with priors on $H_0$, $A_s$ and $n_s$.}

\begin{document}
\maketitle
\flushbottom


\section{Introduction}
\label{sec:intro}

Galaxy and quasar surveys are key observational tools in cosmology. Through them we can 
probe the matter distribution in the Universe and, therefore, test predictions of cosmological 
models and constrain their parameters \cite{Hamilton00,Percival01,Tegmark02,Cole05,
Eisenstein05,Percival07,Sanchez13,Yoon14,Alonso15}. 
A few interesting questions that have been tackled using wide-angle surveys are the homogeneity and isotropy of the 
Universe \cite{Laurent16,Bengaly17,Avila18}, the shape of the distribution function of galaxy density \cite{Clerkin16}, 
the behavior of gravity and the growth of structure 
\cite{Blake11b,Beutler14}, the accelerated expansion of the Universe and the properties of 
dark energy \cite{Anderson12,Carvalho18}, the amount of matter in the Universe \cite{Peacock01}, 
the existence of primordial non-Gaussianities \cite{Ross13} and the mass of neutrinos
\cite{Sorensen12,Loureiro18b}.

To improve constraints on the relevant cosmological parameters -- especially on the equation 
of state of dark energy, the mass of neutrinos, and on non-Gaussianities and other inflation 
signatures -- deeper and wider surveys are desirable \cite{Takada06,Carbone11,Putter17}. 
Moreover, these surveys are required in order to improve our knowledge in various other interesting topics. For instance, the 
cross-correlation between galaxy distribution and gravitational lensing of the cosmic microwave 
background (CMB) -- which serves as a consistency check for our cosmological model, besides 
helping improve constraints on halo bias and on the amplitude of matter perturbations -- peaks 
at large angular scales \cite{Lewis06}. 
This is also the case for the integrated Sachs-Wolfe effect \cite{Ade14}, probed by the cross-correlation 
of CMB temperature maps with galaxy distribution, whose signal comes from the evolution of gravitational potential 
along the CMB photon path. Another 
topic that requires almost full-sky galaxy surveys is the measurement of our own peculiar 
velocity with respect to all other galaxies \cite{Blake02,Erdogdu06,Bilicki11}. Large galaxy surveys can also help us 
improve our understanding of galaxy bias, that could be enhanced at the largest scales if we assume 
that galaxies form at density peaks (see e.g. \cite{Durrer03}).

This type of studies are among the main aims of the largest forthcoming surveys. The Large Synoptic Survey 
Telescope\footnote{\url{http://www.lsst.org/}}
(LSST) \cite{LSST09} will cover $18,000$ $\mathrm{deg^2}$ up to redshifts $z\lesssim 3$, while 
the Euclid space telescope\footnote{\url{http://www.euclid-ec.org}} will observe 15,000 $\mathrm{deg^2}$ and 
detect galaxies up to $z\lesssim 2$ \cite{Amendola13}. The huge volume probed by these surveys 
will lead to such a high number of observed galaxies and quasars that spectroscopic measurements will be 
performed only for a small fraction of them, while for most sources we will have to rely on photometric redshifts 
(photo-$z$s hereafter) and photometric classification. These techniques are, however, plagued with high contamination 
levels, large systematic effects and uncertainties, which must be controlled and understood in order to fully 
mine the relevant datasets.

An existing testbed for such surveys is the WISE$\times$SuperCOSMOS catalog \cite{Bilicki16} (WSC hereafter), a 
full-sky photometric galaxy dataset reaching $z\lesssim 0.4$. Our goal in this paper is to estimate 
cosmological parameters from this sample under the standard cosmological model $\Lambda$CDM. In this process, 
we characterize the systematic effects associated with stellar density (which, although more prominent 
in the WSC, will also be present in LSST and Euclid) and devise a method for mitigating such effects, tested
on simulations and real data. These efforts were aimed at improving the survey's reach toward the largest scales 
($\ell\lesssim20$), commonly distrusted in cosmological analyses (e.g. \cite{Balaguera17,Loureiro18}). 
Finally, we also verified the impact of different assumptions about photo-$z$ properties on the measured parameters. 

This work adds to previous WSC analyses that have investigated its observational and cosmological properties, 
such as hemispherical anisotropies \cite{Bengaly18}, Minkowski functionals \cite{Novaes18}, cross-correlation with CMB lensing 
\cite{Peacock18,Raghunathan18} and temperature \cite{stolzner18}. It also contributes with a new strategy for 
mitigating systematics to those previously proposed \cite{Ross12b,Leistedt13b,Leistedt14,Shafer15,Elsner15,Elsner17}.

The basic technique we employ here -- the measurement of angular power spectra in tomographic redshift shells -- 
has been applied before to other photometric galaxy catalogs such as the MegaZ \cite{Thomas11} 
and the 2MASS Photometric Redshift datasets (2MPZ) \cite{Balaguera17}, to photometric quasar samples 
\cite{Leistedt13b} and spectroscopic galaxy data \cite{Loureiro18}. 

The outline of this paper is as follows:
in Sec.~\ref{sec:dataset} we describe the datasets used in this work, namely the WSC, the Sloan Digital Sky Survey, 
and the Gaia Data Release 2 catalogs. In Secs.~\ref{sec:contamination} and 
\ref{sec:obscuration} we expose the stellar contamination and obscuration effects existent in WSC. 
A method of modeling and mitigating these effects is presented in Sec.~\ref{sec:cleaning}. The WSC data 
had their angular power spectra measured and cosmological parameters inferred using the methodology described 
in Sec.~\ref{sec:methods}. Using simulations, we validate our treatment of systematic effects and our 
measurement methodology in Sec.~\ref{sec:validation}. Our results -- WSC angular power spectra and inferred 
cosmological parameters -- are shown in Sec.~\ref{sec:results}. Sec.~\ref{sec:robustness} presents tests for 
robustness and systematics, and in particular, Sec.~\ref{sec:hemispheres} shows evidence for a systematic effect 
represented by a tension in the cosmological parameters constrained from the angular clustering measured from 
different equatorial hemispheres. Finally, we conclude and summarize our work in Sec.~\ref{sec:conclusions}.


\section{The dataset}
\label{sec:dataset}

The main dataset used in this work  
is the WISE$\times$SuperCOSMOS (WSC) galaxy catalog \cite{Bilicki16}, described 
in Sec.~\ref{sec:wsc}. To estimate the stellar contamination and obscuration in WSC, 
we used the Sloan Digital Sky Survey (SDSS) \cite{York00} and Gaia DR2 \cite{Gaia16,Gaia18} 
photometric catalogs as tracers of stars. These are described in Secs. \ref{sec:sdss} 
and \ref{sec:gaia}, respectively.  


\subsection{WISE$\times$SuperCOSMOS}
\label{sec:wsc}

The WSC catalog is a cross-match between two parent full-sky catalogs: the AllWISE release \cite{Cutri13} 
from the Wide-field Infrared Survey Explorer (WISE) \cite{Wright10}, a mid-infrared space survey in 
four bands $W1$--$W4$ (3.4, 4.6, 12 and 22$\mr{\mu m}$); and the SuperCOSMOS Sky Survey \cite{Hambly01}, 
a program of automated digitalization of optical photographic plates in the $B$, $R$ and $I$ filters, taken by 
the United Kingdom Schmidt Telescope (UKST, in the southern hemisphere) and the Palomar Observatory Sky 
Survey-II (POSS-II, in the northern hemisphere). 

To build the WSC catalog, only the bands $W1$, $W2$, $B$ and $R$ were used. Following the magnitude limits 
of the parent surveys, all WSC sources have magnitudes $W1<17$, $B<21$ and $R<19.5$. To preferentially select 
galaxies over quasars and stars, WSC only includes extended sources according to SuperCOSMOS morphological 
classification \texttt{meanClass}$=1$. It should be noted that the SuperCOSMOS morphological classification
is not as accurate as most recent optical surveys; its angular resolution is approximately $2''$ 
\cite{Hambly01b}, and the quality of the imaging technique is lower (photographic plates vs. CCDs).
 
We also point out that the WSC star/galaxy separation is not based only on SuperCOSMOS morphology but also 
on WISE colours \cite{Bilicki16}. WSC sources have $W1>13.8$ (a cut aimed at removing 
bright stars), $R-W2<7.6-4(W1-W2)$ and $W1-W2<0.9$ (two color-cuts aimed at removing quasars). In this work, 
we further enforced: the requirement $W1-W2>0.2$ for all sources to reduce stellar contamination and maintain 
a constant galaxy selection function across the sky (in opposition to a cut dependent on Galactic latitude 
\cite{Bilicki16}); maximum Galactic extinction, $E(B-V)<0.10$; removal of sources in highly contaminated 
regions such as near the Galactic plane and bulge, around the Magellanic Clouds and in regions presenting unusually high 
densities according to a lognormal distribution, all accomplished with the WSC final mask \cite{Bilicki16}; 
removal of sources in regions where Gaia stellar density is greater than 7 times its average (this increases 
the masked region around the bulge; see Sec.~\ref{sec:gaia}); and removal of a SuperCOSMOS tile with bad photometry.

\begin{figure}
  \center
  \includegraphics[width=1.0\textwidth]{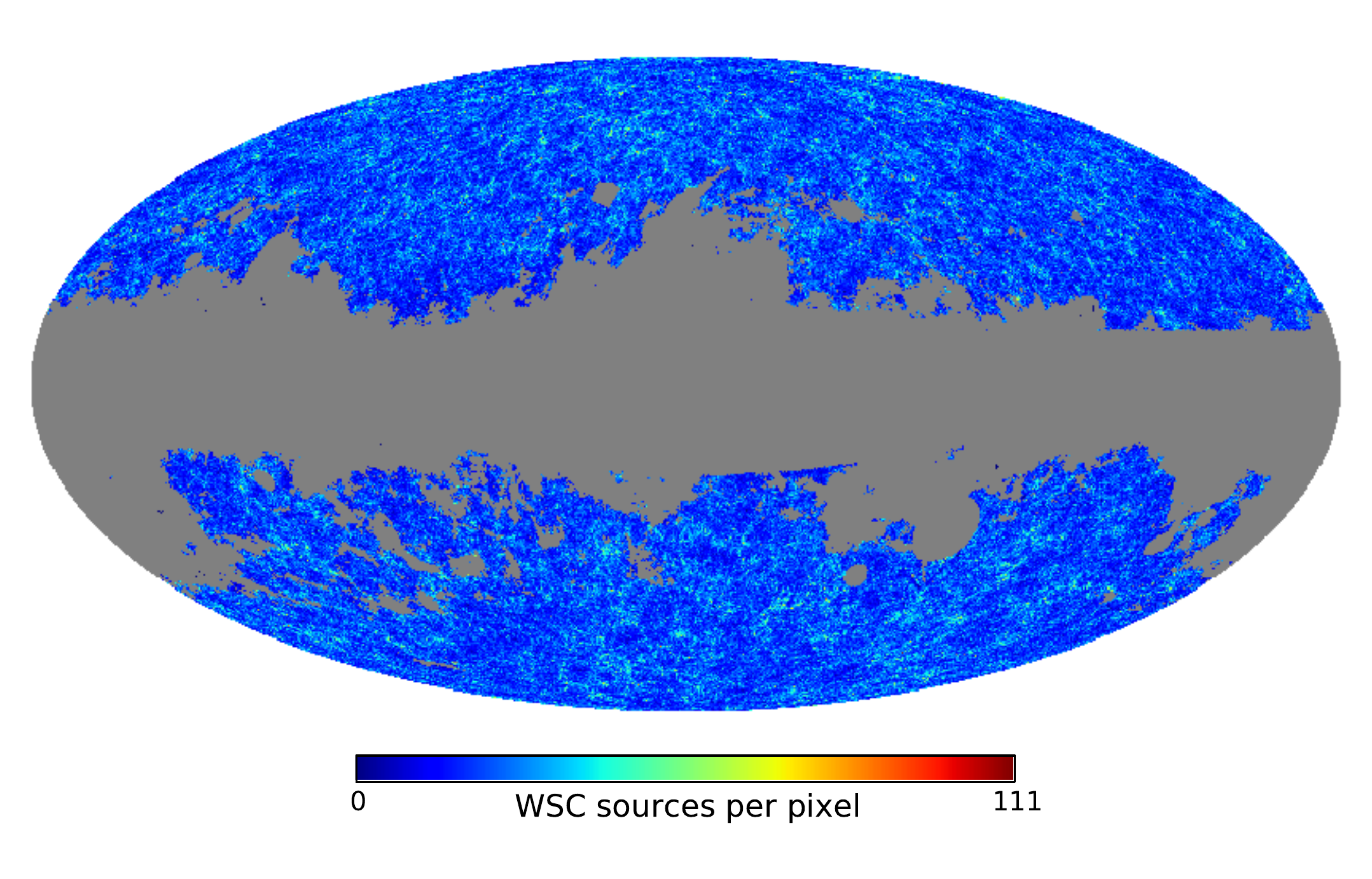}
  \caption{Sky map of number counts of WSC sources (mostly galaxies) in the photo-$z$ range $0.10<z<0.35$ in Galactic coordinates, 
    after applying all the cuts described in Sec.~\ref{sec:dataset}, using Healpix \cite{Gorski05} resolution 
    parameter $N_{\mr{side}}=256$. The gray regions were masked out.}
  \label{fig:wsc-map}
\end{figure}

\begin{figure}
  \center
  \includegraphics[width=0.7\textwidth]{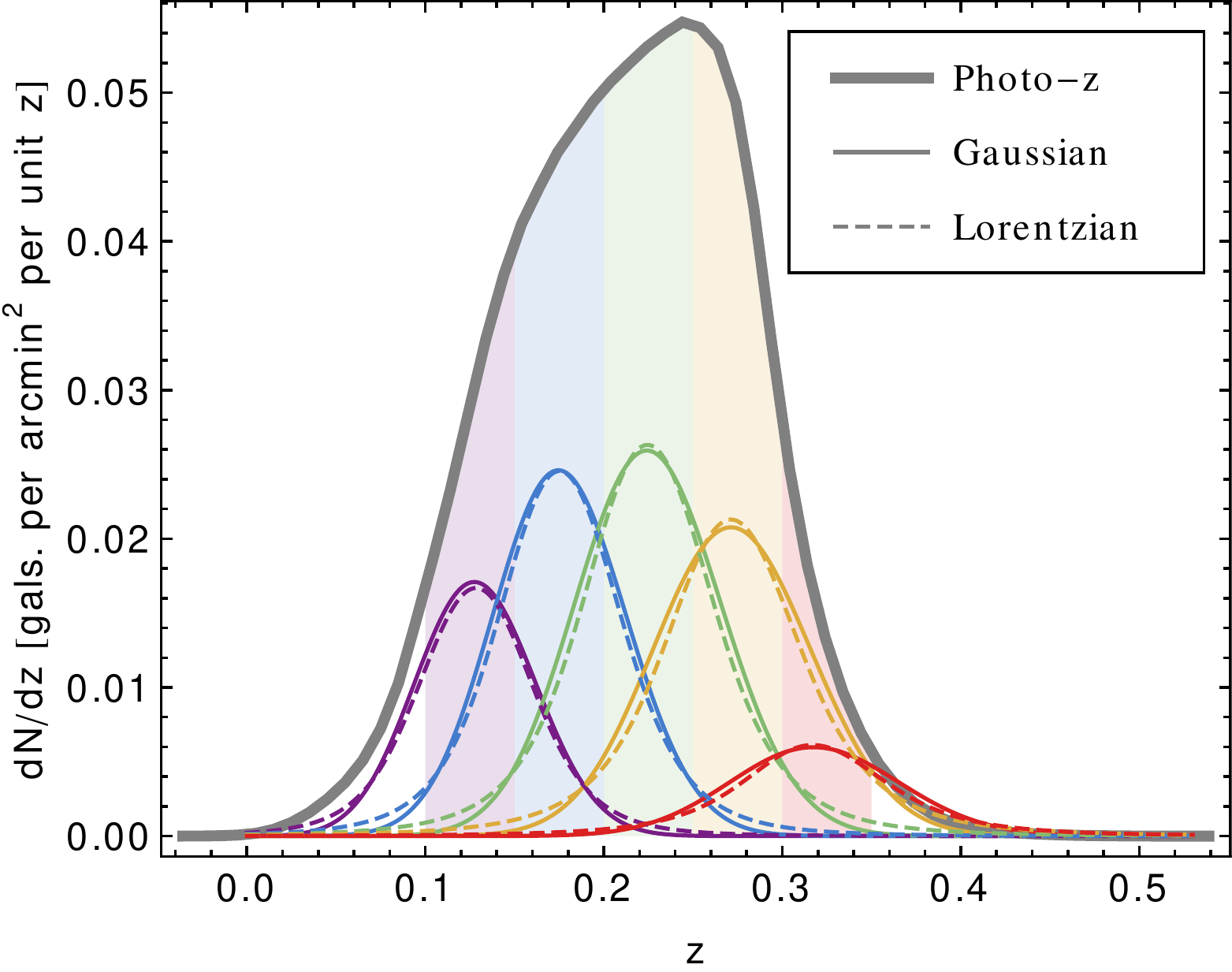}
  \caption{Redshift distribution of WSC sources, given in number of galaxies per arcmin$^2$, per unit redshift. 
The thick gray line represents their distribution in terms of their photo-$z$s, after all cuts described in 
Sec.~\ref{sec:dataset} have been applied. The thin, colored lines represent their estimated spectroscopic redshift 
distributions inside each photo-$z$ bin (represented as colored bands), 
    assuming Gaussian (solid lines) and generalized Lorentzian (dashed lines) photo-$z$ errors, both computed with 
    Eq.~(\ref{eq:spec-z}).}
  \label{fig:wsc-z-dist}
\end{figure}

The photo-$z$s for all the WSC sources were obtained with the artificial neural network code 
ANN$z$ \cite{Collister04}\footnote{\url{http://www.homepages.ucl.ac.uk/~ucapola/annz.html}}, trained 
on a WSC cross-match with the complete and deep Galaxy And Mass Assembly (GAMA)-II spectroscopic dataset 
\cite{Driver09}. The adequacy of GAMA for photo-$z$ training in WSC has been discussed
and demonstrated in \cite{Bilicki16}. As explained there, these photo-$z$s were further corrected for asymmetries 
between their northern and southern distributions, likely caused by differences between the POSS-II and UKST 
pass-bands. That paper contains also a comprehensive analysis of photo-$z$
properties based on comparisons with several external spectroscopic datasets.
A projection of all sources in the photometric redshift (photo-$z$) range $0.10<z<0.35$ is presented 
in Fig.~\ref{fig:wsc-map}, and Fig.~\ref{fig:wsc-z-dist} shows the photo-$z$ distribution of WSC galaxies. 
In this work we binned the WSC sources in photo-$z$ shells of width $\Delta z = 0.05$, covering the range $0.10<z<0.35$. 
However, for reasons presented in Sec.~\ref{sec:character}, most of our analysis ignores 
the first bin ($0.10<z<0.15$).

It was shown in \cite{Peacock18} that the WSC photo-$z$ errors can be 
modeled by a Gaussian with zero mean (i.e. no bias) and standard deviation 
$\sigma(z_{\mr{p}})=0.02+0.08z_{\mr{p}}$. Also, that an even better model would be a generalized Lorentzian, where
the probability density of a galaxy with photo-$z$ $z_{\mr{p}}$ to have true redshift $z_{\mr{s}}$ is:
\begin{equation}
p(z_{\mr{s}}|z_{\mr{p}}) = \frac{ \sqrt{2\pi a(z_{\mr{p}})} \Gamma\left[a(z_{\mr{p}})-\frac{1}{2}\right] }
{ \Gamma\left[a(z_{\mr{p}})\right] } 
\left[ 1+\frac{(z_{\mr{s}}-z_{\mr{p}})^2} {2a(z_{\mr{p}})s^2(z_{\mr{p}})} \right]^{-a(z_{\mr{p}})},
\label{eq:lorentzian}
\end{equation} 
where $\Gamma(x)$ denotes the Gamma function. The WSC photo-$z$s follows $a(z)=3-4z$ and 
$s(z)=0.02+0.04z$.  

The spectroscopic redshift (spec-$z$) distribution $\mr{d}N^i/\mr{d}z_{\mr{s}}$ of the galaxies in our photo-$z$ bin $i$ 
can be estimated from a convolution of $p(z_{\mr{s}}|z_{\mr{p}})$ and the photo-$z$ distribution $\mr{d}N/\mr{d}z_{\mr{p}}$:
\begin{equation}
\frac{\mr{d}N^i}{\mr{d}z_{\mr{s}}} = \int_{z^i_{\mr{min}}}^{z^i_{\mr{max}}} p(z_{\mr{s}}|z_{\mr{p}}) \frac{\mr{d}N}{\mr{d}z_{\mr{p}}} \mr{d}z_{\mr{p}},
\label{eq:spec-z}
\end{equation} 
where the $i$th bin in photo-$z$ ranges from $z_{\mr{min}}^i$ to $z_{\mr{max}}^i$. The resulting distributions in the different redshift bins are 
shown in Fig.~\ref{fig:wsc-z-dist}. We verified that the expected spec-$z$ distributions in each bin are very well 
fitted by a Gaussian (generalized Lorentzian) when the photo-$z$ errors are Gaussian (generalized Lorentzian).
Also, one can see in Fig.~\ref{fig:wsc-z-dist} that the mean values of the spec-$z$s in every bin are practically independent of the 
distribution assumed for the photo-$z$ errors, whereas the FWHM may depend on it. Thus, when 
computing theoretical angular power spectra $C_\ell$s to fit the data, we decided to keep the means of the spec-$z$ 
distributions fixed and let the widths vary as nuisance parameters. 


\subsection{Sloan Digital Sky Survey}
\label{sec:sdss}

The Sloan Digital Sky Survey (SDSS, \cite{York00}) is one of the largest astronomical surveys in operation. 
Here we use its $14^{\mr{th}}$ Data Release (DR14) \citep{Abolfathi18}. 
The survey has already covered $\sim 1/3$ of the northern hemisphere, imaging in 
the five $ugriz$ broad bands. The effective photometric limit of the survey is $r=22.2$ ($95\%$ completeness 
for point sources). In addition, its spectroscopic counterpart is complete down to $r=17.77$ 
for the Main Galaxy Sample (MGS) and deeper for other sub-projects, such as the Baryon Oscillation Spectroscopic 
Survey (BOSS) \citep{Anderson12}. 

Despite containing a wealth of cosmological information, the SDSS data was only used here to verify the 
existence of stellar contamination and obscuration in the WSC catalog, as described in Sec.~\ref{sec:character}.
The SDSS star-galaxy separation classifies extended objects as galaxies and those point-like as stars. 
This classification is based on a difference between two types of magnitudes, namely sources with
$\mathrm{psfMag - cModelMag} > 0.145$ are classified as galaxies and otherwise as 
stars.\footnote{The two types of magnitudes $\mathrm{psfMag}$ and $\mathrm{cModelMag}$ represent 
magnitudes obtained using the point spread function (PSF) model and the best linear combination of 
exponential and de Vaucouleurs profiles, respectively. Note that the `stars' according to this definition 
will include any point sources, also extragalactic ones such as quasars.} The PSF of $1.24$ arcsec (full width at half maximum, FWHM) 
allows for a reliable star-galaxy separation up to $r<21.5$ \cite{Lupton02} 
and accurate astrometric positions ($<0.1$ arcsec per coordinate) at $r<20.5$ \cite{Pier03}.


\subsection{Gaia}
\label{sec:gaia}

Gaia is a space mission of the European Space Agency (ESA) aimed at measuring the three-dimensional positions 
and velocities of about $10^9$ Milky Way stars \cite{Gaia16}. It uses two optical telescopes of $0.7 \mr{m^2}$ 
collecting area equipped with astrometry, photometry and spectrometry instruments. The astrometry is performed 
in the range  $330-1050$ nm ($G$-band hereafter). 
Gaia Data Release 2 (DR2, \cite{Gaia18}) contains $1.3\times 10^9$ point sources with measured parallaxes and proper motions at  
$\sim 1$ milliarcsecond (mas) and $\sim 1$ mas per year ($\mr{mas~y^{-1}}$) precision, respectively, up to a 
$G$ band magnitude of 20.

\begin{figure}
  \center
  \includegraphics[width=1.0\textwidth]{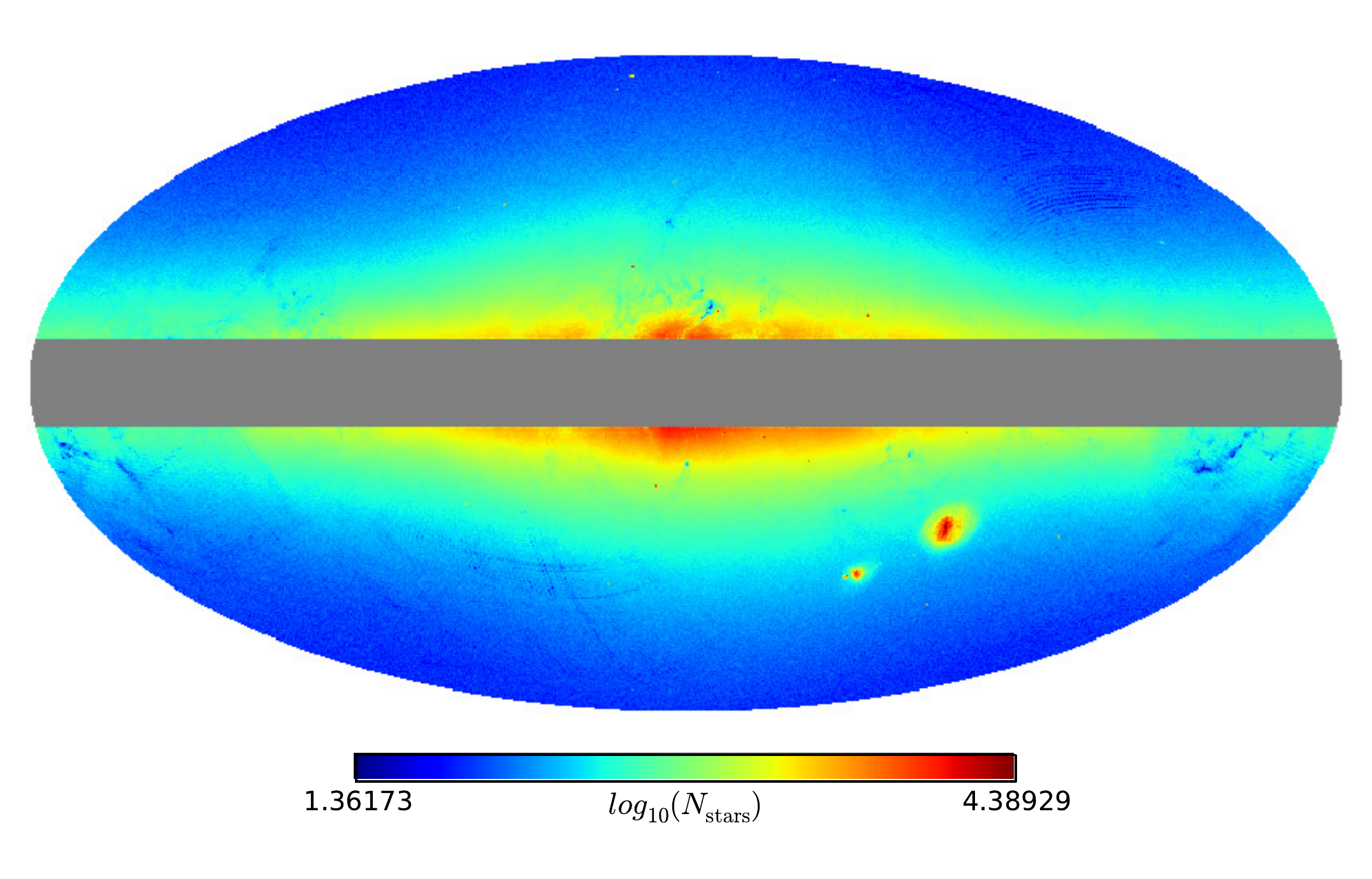}
  \caption{Sky map of number counts (sources per pixel, using $N_{\mr{side}}=256$, in logarithmic scale) at Galactic latitudes 
    $|b|>10^\circ$ of Gaia DR2 sources at $G<20$ magnitude limit, after removing those
    with both parallax and proper motion measurements consistent with zero.}
  \label{fig:gaia-map}
\end{figure}

In order to create a representative full-sky stellar density map using the Gaia DR2 source catalog, 
we had to ensure minimal observational effects (i.e. anisotropic completeness) and minimal contamination 
by galaxies and quasars. To achieve the first goal we applied a magnitude cut $G<20$; as the limiting 
magnitudes in the regions of interest (away from the Galactic plane) are $G\gtrsim 20.8$ \cite{Arenou18}, 
we expect uniform sky completeness under this selection criterion. The second goal was achieved by
taking advantage of the excellent astrometric measurements of Gaia: we assumed that the majority of 
observed stars should have either a non-zero parallax or a non-zero total proper motion. Thus, we enforced,
for all the selected sources, a 5$\sigma$ detection of at least one of these quantities. We computed the total 
proper motion as $\mu = \sqrt{\mu_\delta^2 + \mu_{\alpha'}^2}$,
where $\mu_{\alpha'}\equiv\mu_\alpha\cos\delta$, $\mu_\alpha$ is the proper motion in the direction of right ascension 
$\alpha$ and $\mu_\delta$ is the proper motion in the direction of declination $\delta$. The 
uncertainty on $\mu$ was computed with the usual error propagation, taking into account the 
covariance between $\mu_\delta$ and $\mu_{\alpha'}$ measurement errors.

In order to avoid regions of high stellar density, here we only selected sources at Galactic latitudes $|b|>10^\circ$. A map of the $282,045,470$ selected Gaia sources, in Galactic coordinates, is presented in 
Fig.~\ref{fig:gaia-map}. 


\section{Characterizing and cleaning the WSC catalog}
\label{sec:character}

\subsection{Stellar contamination}
\label{sec:contamination}

As described in \cite{Bilicki16}, the WSC catalog is affected by considerable stellar contamination, especially 
near the Galactic plane. In order to demonstrate this contamination, we follow the strategy applied in 
\cite{Bilicki16} of cross-matching WSC with SDSS photometric sources, taking advantage of the 
latter's better star-galaxy separation. Our final analysis, however, is based on a stellar density template 
derived from Gaia's observations (see Sec. \ref{sec:cleaning}). 

\begin{figure}
  \center
  \includegraphics[width=0.65\textwidth]{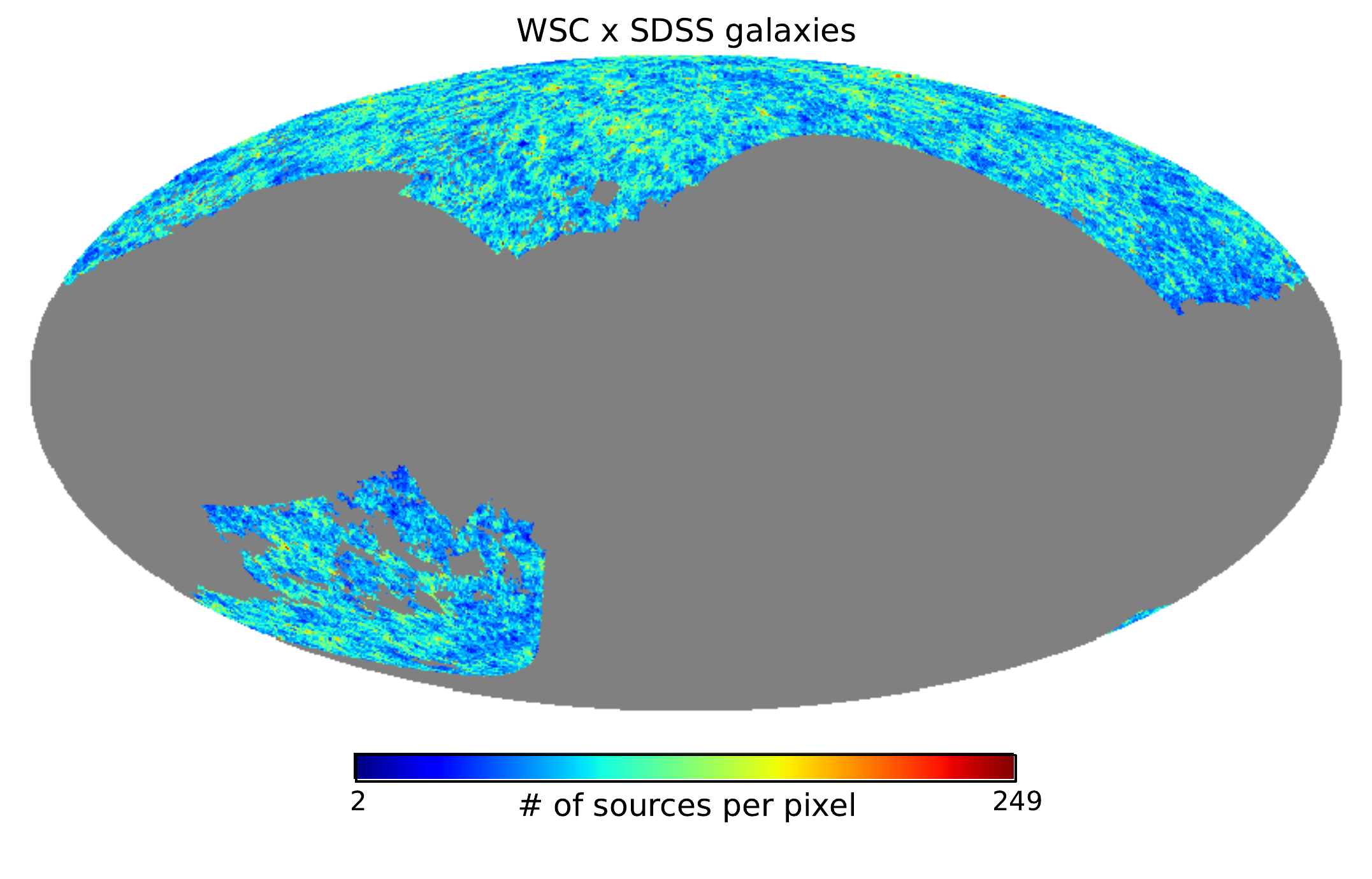}
  \includegraphics[width=0.65\textwidth]{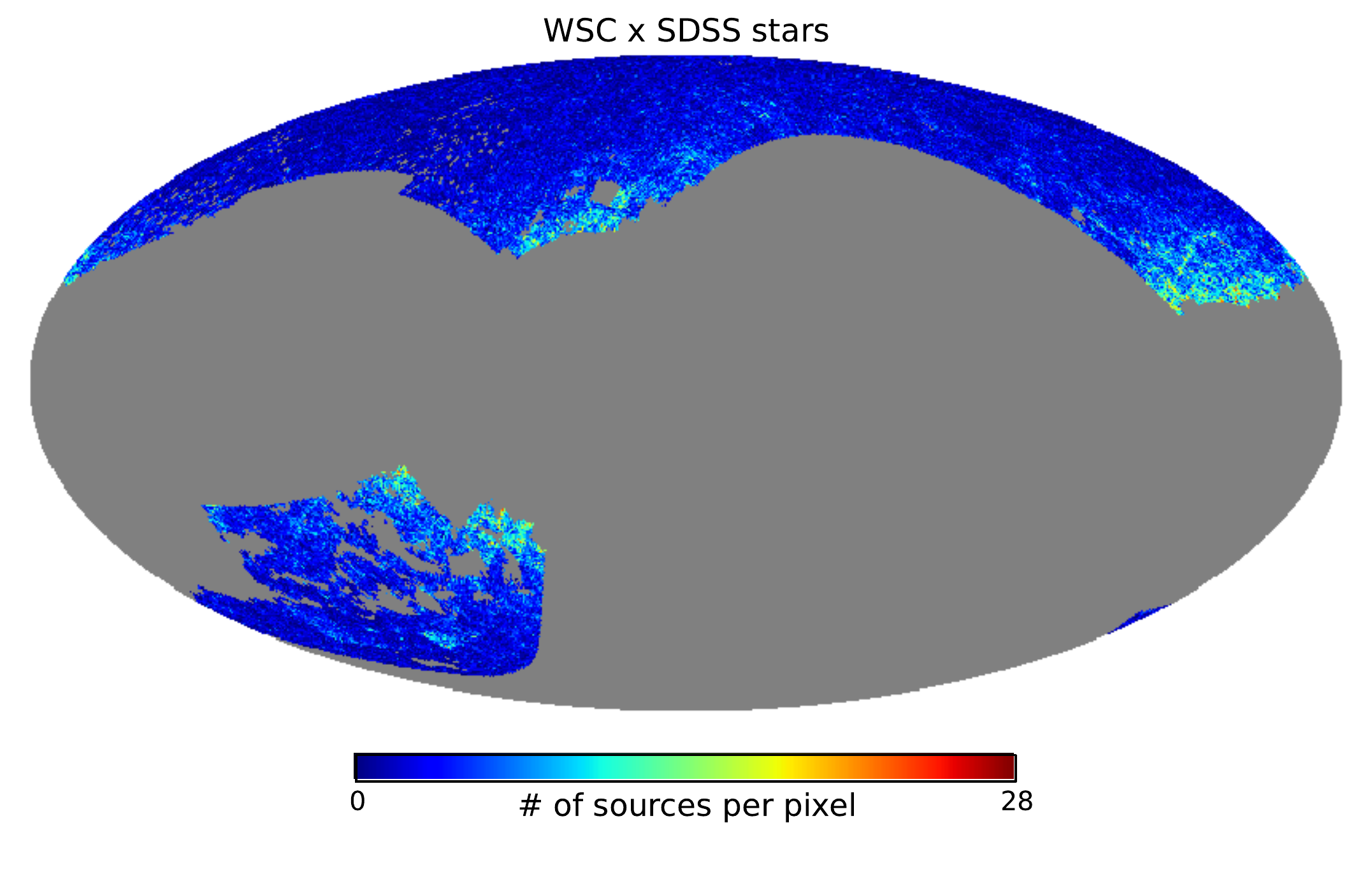}
  \caption{Number counts map in Galactic coordinates of the cross-match between SDSS and 
    our WSC sources, selected according to Sec.~\ref{sec:wsc}. In this particular case we used 
      $N_{\mr{side}}=128$ to improve visualization. 
    The top (bottom) panel shows the cross-match for the SDSS sources classified as galaxies (stars).}
  \label{fig:wscXsdss-map}
\end{figure}

We cross-matched the WSC sample and the photometric SDSS DR14, 
linking the nearest objects in the latter (within 2 arcsec) to objects in the former. In total, our cross-matched 
sample consists of $4,282,564$ sources which includes both stars and galaxies.
The result of this cross-match is shown in Fig.~\ref{fig:wscXsdss-map}. From the bottom panel, it is quite clear 
that stellar contamination exists and increases towards the Galactic plane and bulge. Also, there are hints 
that the WSC galaxies are partly obscured by stars, since the top panel shows a deficit of galaxies in the same 
regions where stars creep in. Overall, 96\% percent of the matched sources are galaxies and 4\% are stars, 
according to SDSS. 

\begin{figure}
  \center
  \includegraphics[width=0.7\textwidth]{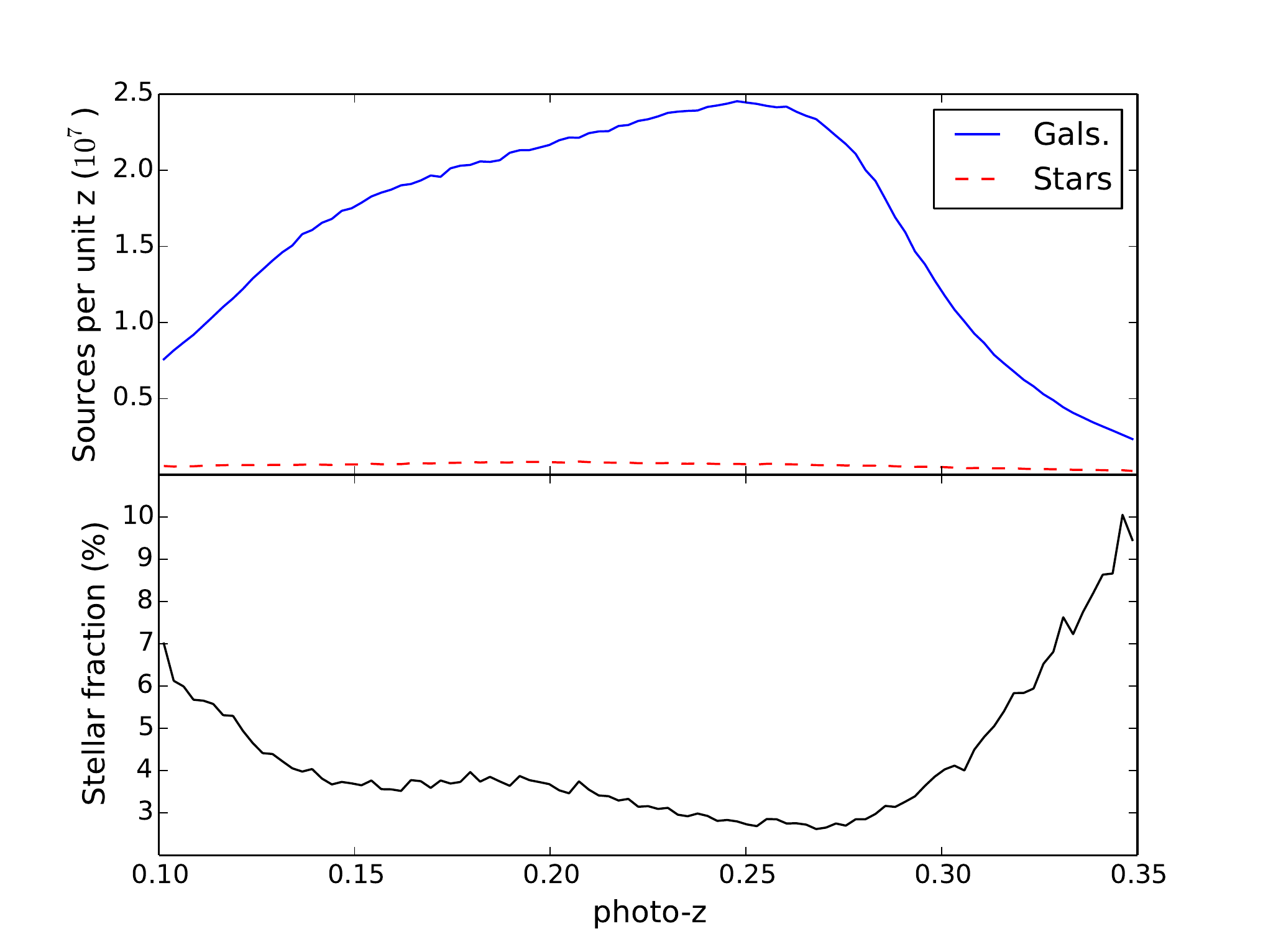}
  \caption{\emph{Top:} total number (per unit redshift) of WSC sources matched with 
    SDSS galaxies (blue solid line) and stars (red dashed line). 
    \emph{Bottom:} fraction of matched sources that are stars according to SDSS, 
    as a function of WSC photo-$z$.}
  \label{fig:wscXsdss-star-fraction}
\end{figure}

Fig.~\ref{fig:wscXsdss-star-fraction} shows how the amount of stellar contamination changes with 
redshift. We can see that, at high redshifts, the contamination (i.e. stellar fraction) increases 
significantly. This increase (and the one at low redshift) are mostly caused by a drop in the 
number of observed galaxies and not by an increase in the number of misclassified stars (see the top panel).

\begin{figure}
  \center
  \includegraphics[width=0.9\textwidth]{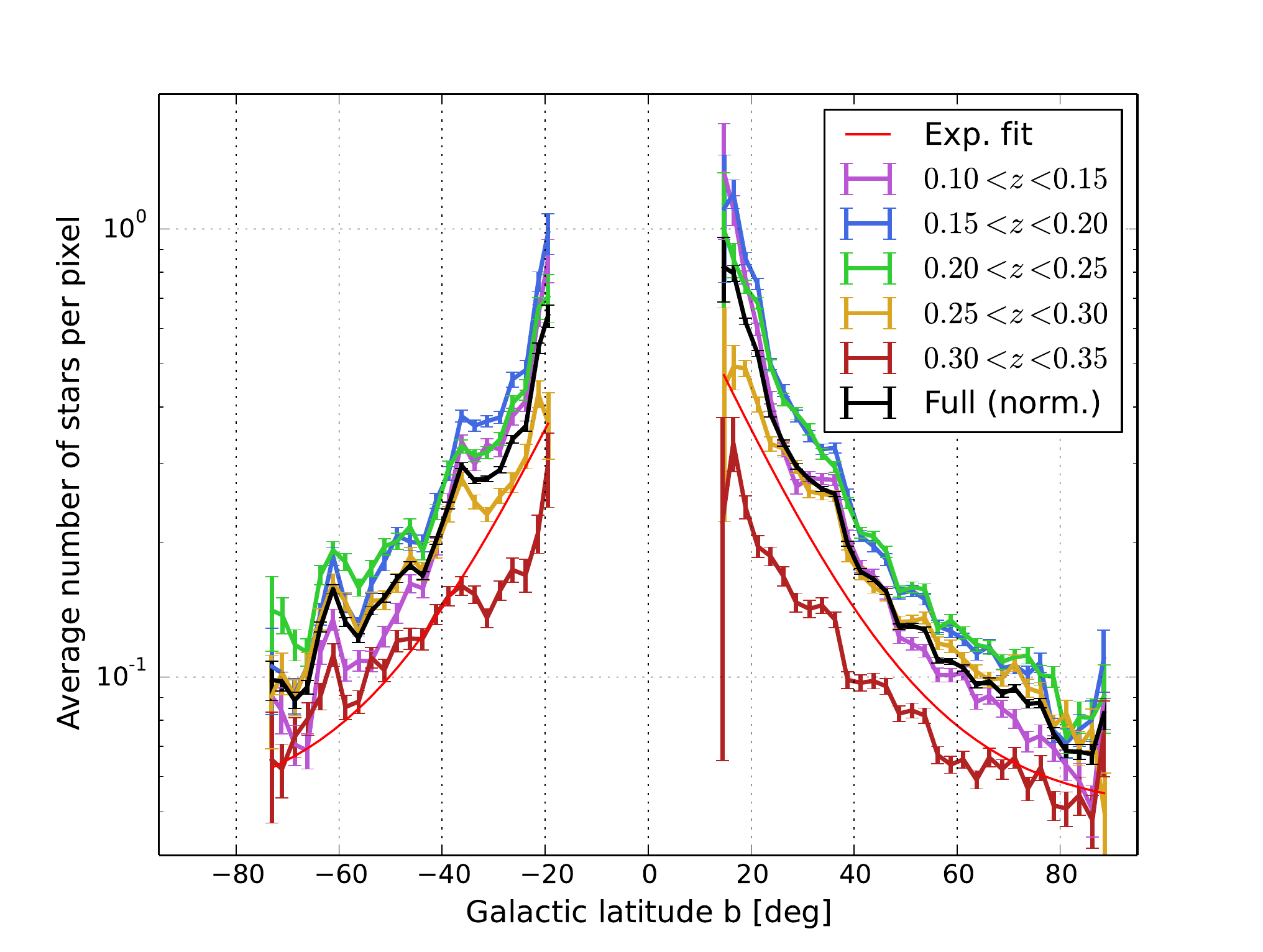}
  \caption{Average number of WSC sources matched to SDSS stars per pixel ($N_{\mr{side}}=256$) as a function of 
  Galactic latitude $b$. Each thick colored curve (with error bars) represents a different photo-$z$ bin, and 
  the black curve represents the full range $0.10<z<0.35$, normalized by 5 to ease comparison with 
  the other curves. The error bars represent Poisson noise inside the $b$ bins used to build the plot. The thin 
  red line shows the result of an exponential fit (Eq.~\ref{eq:exp-n-stars}) to the black line, scaled by 0.71 
  to unclutter the plot.}
  \label{fig:wscXsdss-star-b}
\end{figure}

The average dependence with Galactic latitude $b$ of the number of stars contaminating WSC 
(according to SDSS classification) is shown in Fig.~\ref{fig:wscXsdss-star-b}. We see that 
both Galactic hemispheres, as well as different redshift ranges, follow very similar 
trends. For instance, the binned redshift curves differ from the full range one mostly by a 
constant factor. Despite these similarities, subtle slope changes and hemispherical asymmetries
are noticeable. 

Figure \ref{fig:wscXsdss-star-b} also shows the result of an exponential fit to the average number of stars 
per pixel $n_{\mr{stars}}$ in the range $0.10<z<0.35$ as a function of Galactic latitude $b$:
\begin{equation}
n_{\mr{stars}}(b) = \bar{n}e^{-a |b|}+c,
\label{eq:exp-n-stars}
\end{equation}
where $\bar{n}=7.14$, $a=0.06$ and $c=0.35$; in the plot, the fit was scaled by 0.71 to ease visualization.
We can see that the fit represents the stellar density reasonably well.

\subsection{Stellar obscuration}
\label{sec:obscuration}

As shown in Fig.~\ref{fig:wscXsdss-map} and already noted in previous works with SDSS data \cite{Ross12b}, the density of observed galaxies anti-correlates with the stellar density: regions with high concentration 
of stars tend to present lower galaxy counts. We call this phenomenon \emph{stellar obscuration}. 
This is expected to affect mostly photometric datasets, such as WSC, due to issues such as 
blending and source misidentification. Spectroscopic surveys can avoid stellar contamination given 
that spectroscopy can easily separate stars from galaxies, but even they suffer from such obscuration, 
as shown in \cite{Ross12b}.

The approach used in \cite{Ross12b} to estimate stellar obscuration was to correlate the observed number of 
galaxies with the density of foreground stars: as galaxies are, on average, homogeneously distributed on the sky, these two should be 
uncorrelated in the absence of systematic effects. This approach is feasible in the spectroscopic SDSS case 
because the galaxy sample there is free from stellar contamination. For WSC sources that include a large amount of 
stars disguised as galaxies, however, this approach is not possible: correlating WSC sources with stellar density 
would mix contributions from obscuration and contamination. 

An alternative approach is to look at the standard deviation of the counts of galaxies inside pixels as a function of stellar
density. The basic idea is that stellar obscuration will not only modulate the mean observed density 
$\bar{n}_{\mr{obs}}$ but also the cosmological fluctuations around the mean, since the former multiplies the latter.
Let us formalize this statement through a model for the observed projected density of sources 
$n^i_{\mr{obs}}(\bm{\theta})$ at angular position $\bm{\theta}$, for a particular redshift bin $i$, that takes into 
account both contamination and obscuration:
\begin{equation}
n^i_{\mr{obs}}(\bm{\theta}) = W(\bm{\theta})\{[1-\alpha_i S(\bm{\theta})]n^i_{\mr{g}}(\bm{\theta}) +
\beta_i S(\bm{\theta}) + \epsilon_{\mr{g},i}(\bm{\theta}) + \epsilon_{\mr{s},i}(\bm{\theta})\}.
\label{eq:sources-model}
\end{equation}
In the equation above, $W(\bm{\theta})$ is the survey binary window function, $S(\bm{\theta})$ is a 
template for the expected density of stars, $n^i_{\mr{g}}(\bm{\theta})$ is the real galaxy density, 
and $\epsilon_{\mr{g},i}(\bm{\theta})$ and $\epsilon_{\mr{s},i}(\bm{\theta})$ are Poisson fluctuations 
in the number of observed galaxies and stars, respectively. The non-negative parameters $\alpha_i$ and $\beta_i$ 
control the amount of obscuration and contamination in the $i$th bin, respectively. The noise terms obey the statistical 
properties $\langle\epsilon_{\mr{g},i}(\bm{\theta})\rangle=\langle\epsilon_{\mr{s},i}(\bm{\theta})\rangle=0$, 
$\langle\epsilon_{\mr{g},i}(\bm{\theta})\epsilon_{\mr{s},i}(\bm{\theta})\rangle=0$,
$\langle\epsilon^2_{\mr{g},i}(\bm{\theta})\rangle=W(\bm{\theta})[1-\alpha_i S(\bm{\theta})]\bar{n}_{\mr{g},i}$ and 
$\langle\epsilon_{\mr{s},i}^2(\bm{\theta})\rangle=W(\bm{\theta})\beta_i S(\bm{\theta})$, such that 
the variance of the total noise is equal to the expected number of observed sources, 
$\langle[\epsilon_{\mr{g},i}(\bm{\theta})+\epsilon_{\mr{s},i}(\bm{\theta})]^2\rangle=\langle n^i_{\mr{obs}}(\bm{\theta})\rangle$.

Our model for the observed sources (Eq.~\ref{eq:sources-model}) makes three important simplifications: first, the number of 
observed galaxies is linearly suppressed by the stellar density; second, both contamination and obscuration 
depend on the same template; third, the template $S(\bm{\theta})$ is independent of redshift, 
although the final obscuration and contamination are modulated by bin-dependent parameters $\alpha_i$ and 
$\beta_i$. The fact that both obscuration and contamination follow the same template leads 
to a replacement of galaxies for stars, and thus we call this combined effect ``usurper contamination''.

To estimate $\alpha_i$ from the observed variance of $n^i_{\mr{obs}}$, we first write $\bm{\theta}$ as
a pair of generalized coordinates $\bm{\theta}=(s,\lambda)$, where $s$ specifies an isocontour of 
$S(\bm{\theta})$ while by varying $\lambda$ we move along this isocontour. We then compute 
the mean of $n^i_{\mr{obs}}$ for a fixed $s$:

\begin{equation}
\bar{n}^i_{\mr{obs}}(s) = \frac{1}{A(s)} \int n^i_{\mr{obs}}(s,\lambda) \mr{d}\lambda = 
[1-\alpha_iS(s)]\bar{n}_{\mr{g},i}(s) + \beta_i S(s),
\label{eq:iso-mean}
\end{equation}
where
\begin{equation}
A(s) \equiv \int W(s,\lambda) \mr{d}\lambda,
\label{eq:iso-area}
\end{equation}
and
\begin{equation}
\bar{n}_{\mr{g},i}(s) \equiv \frac{1}{A(s)} \int W(s,\lambda) n_{\mr{g}}^i(s,\lambda) \mr{d}\lambda.
\label{eq:iso-gals}
\end{equation}
Note that due to our choice of coordinates, the value of $S(\bm{\theta})$ only depends on $s$. Also, 
we assumed that when averaging the noise $\epsilon_{\mr{g},i}+\epsilon_{\mr{s},i}$ along $\lambda$, we may 
approximate the result to zero. Now we can define the difference 
$\delta n_{\mr{obs},i}(s,\lambda) \equiv n^i_{\mr{obs}}(s,\lambda) - W(s,\lambda) \bar{n}^i_{\mr{obs}}(s)$ and 
compute the variance of source number counts for an isocontour $s$ of $S(\bm{\theta})$:

\begin{equation}
\sigma_{\mr{obs},i}^2(s) = \frac{1}{A(s)} \int \delta^2_{\mr{obs},i}(s,\lambda) \mr{d}\lambda \simeq
[1-\alpha_i S(s)]^2v^i_{\mr{g}} + \bar{n}_{\mr{obs}}^i(s),
\label{eq:iso-var}
\end{equation}
where $v^i_{\mr{g}}=\langle[n^i_{\mr{g}}(s,\lambda)-\bar{n}_{\mr{g},i}(s)]^2\rangle$ is the true galaxy variance 
(i.e. it does not include systematic effects nor shot-noise). Eq.~(\ref{eq:iso-var}) assumes that 
the mean source number counts in $s$, $\bar{n}_{\mr{obs}}^i(s)$, approximates the expected value 
$\langle n^i_{\mr{obs}}(s,\lambda)\rangle$, and that $v^i_{\mr{g}}$ is independent of the angular position. 
By placing all directly measurable quantities on the left hand side, we get a function of $s$ that 
might be fitted to extract the obscuration parameter $\alpha_i$:

\begin{equation}
\sigma_{\mr{obs},i}^2(s) - \bar{n}_{\mr{obs}}^i(s) = [1-\alpha_i S(s)]^2v^i_{\mr{g}}.
\label{eq:iso-model}
\end{equation}
In the absence of obscuration, the quantity on the left would be the variance due to cosmological fluctuations 
(and thus we call it $\sigma_{\mr{cosmo}}^2$) and it should be constant all over the sky.
We point out that, in practice, the estimation of $\sigma_{\mr{cosmo}}^2$ is performed in a discrete way [the integral in 
Eq.~(\ref{eq:iso-var}), for instance, is actually a sum over pixels], and the isocontours are approximated 
by bands of similar $S(\bm{\theta})$ values.

\begin{figure}
  \center
  \includegraphics[width=1.0\textwidth]{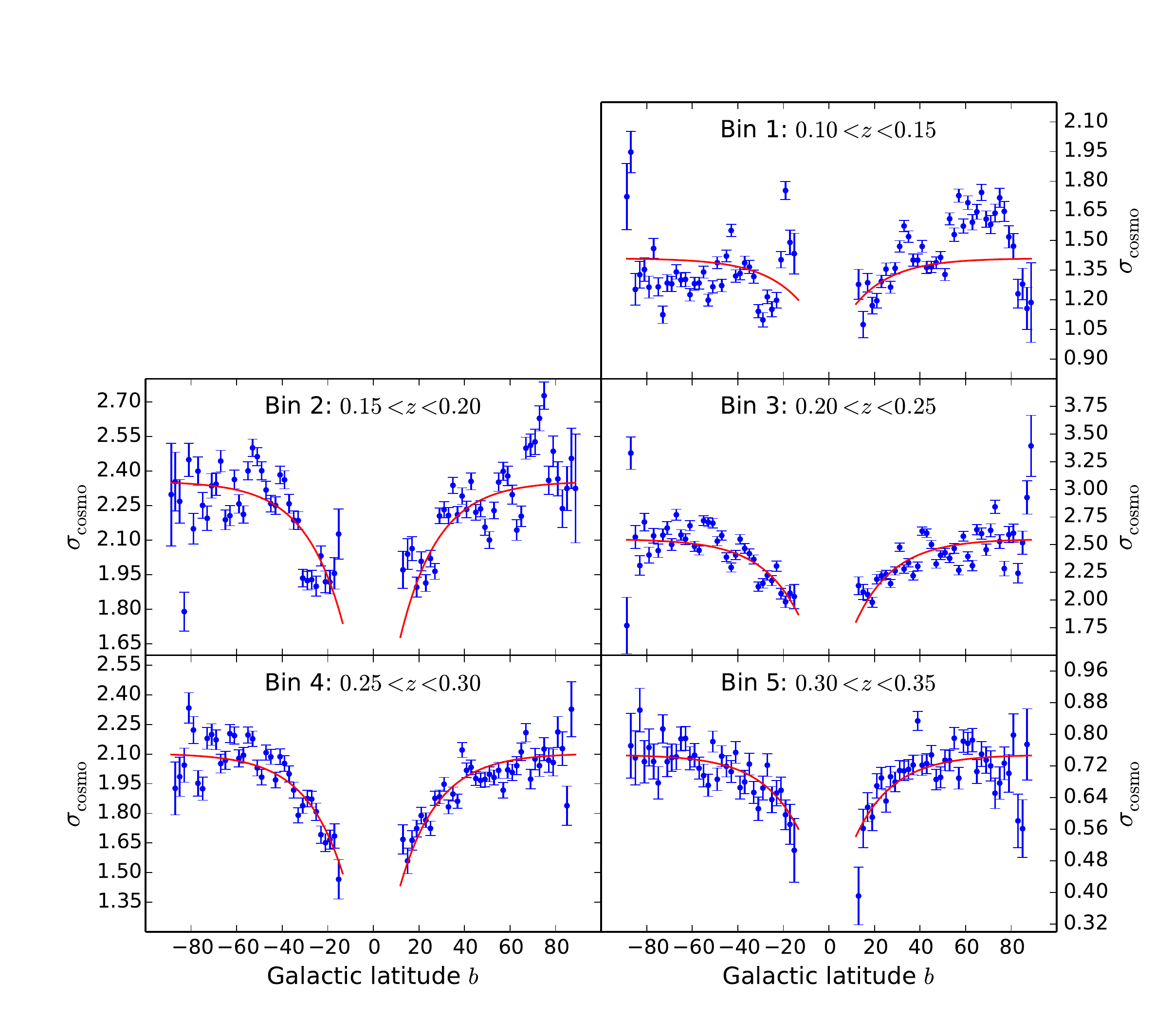}
  \caption{Each panel shows, for a different redshift bin, the cosmological contribution to the 
    standard deviation of WSC number counts in pixels ($N_{\mr{side}}=256$), computed inside bands 
    of fixed Galactic latitude $b$ (blue data points). The red curve is the fitted model given by the 
    square root of Eq.~(\ref{eq:iso-model}), with stellar template $S(\bm{\theta})$ given by 
    Eq.~(\ref{eq:exp-n-stars}) (fixed according the contamination estimated in Sec.~
    \ref{sec:contamination}).}
  \label{fig:wsc-dev-b}
\end{figure}

Fig.~\ref{fig:wsc-dev-b} shows $\sigma_{\mr{cosmo}}$ as a function of Galactic latitude $b$ for each photo-$z$ bin. 
It is completely evident that the amplitude of the 
observed fluctuations is not constant and decreases towards the Galactic plane, quantitatively proving the 
existence of obscuration in the WSC data. This figure also shows that our obscuration model, given by the 
square root of Eq.~(\ref{eq:iso-model}) (depicted as a red line) describes really well the overall behavior
of this effect (except for the first bin). Here it is important to emphasize that only $\alpha_i$ 
and $v^i_{\mr{g}}$ were fitted to $\sigma_{\mr{cosmo}}$, while $S(s)$ was held fixed according to 
Eq.~(\ref{eq:exp-n-stars}), estimated from contamination. Thus, Fig.~\ref{fig:wsc-dev-b} also shows that our 
assumption that both obscuration and contamination follow the same template $S(s)$ is reasonable.

The error bars in Fig.~\ref{fig:wsc-dev-b} were computed by error propagation from the uncertainties on 
$\bar{n}_{\mr{obs}}^i(s)$ and $\sigma_{\mr{obs},i}(s)$, which are given by $\sigma_{\mr{obs},i}(s)/\sqrt{N(s)}$ and 
$\sigma_{\mr{obs},i}(s)/\sqrt{2[N(s)-1]}$, respectively [$N(s)$ is the number of pixels taken into account in each isocontour]. 
We interpret the fact that the scatter of the data points is larger than the one expected from the error bars 
as an indication that $v^i_{\mr{g}}$ actually varies from one isocontour to another due to large-scale 
density fluctuations that are the size of the isocontours. It might also indicate that other factors 
are in play (e.g. Galactic extinction, seeing and bad photographic plates) 
or that the template $S(\bm{\theta})$ can be improved.   

\subsection{Cleaning the WSC projected galaxy distribution}
\label{sec:cleaning}

We can note from Fig.~\ref{fig:wscXsdss-star-b} that the stellar contamination is not perfectly 
symmetric across the Galactic plane. Moreover, Fig.~\ref{fig:gaia-map} shows that the stellar 
density (and likely contamination) is larger near the Galactic bulge, as expected. Aiming at 
improving the removal of contamination and obscuration from WSC projected galaxy distribution, we  
built a stellar density template $S(\bm{\theta})$ from the Gaia DR2 map presented in Fig.~\ref{fig:gaia-map}, 
which should accurately represent the stellar distribution -- and 
thus contamination and obscuration -- all over the sky. 

\begin{figure}
  \center
  \includegraphics[width=1.0\textwidth]{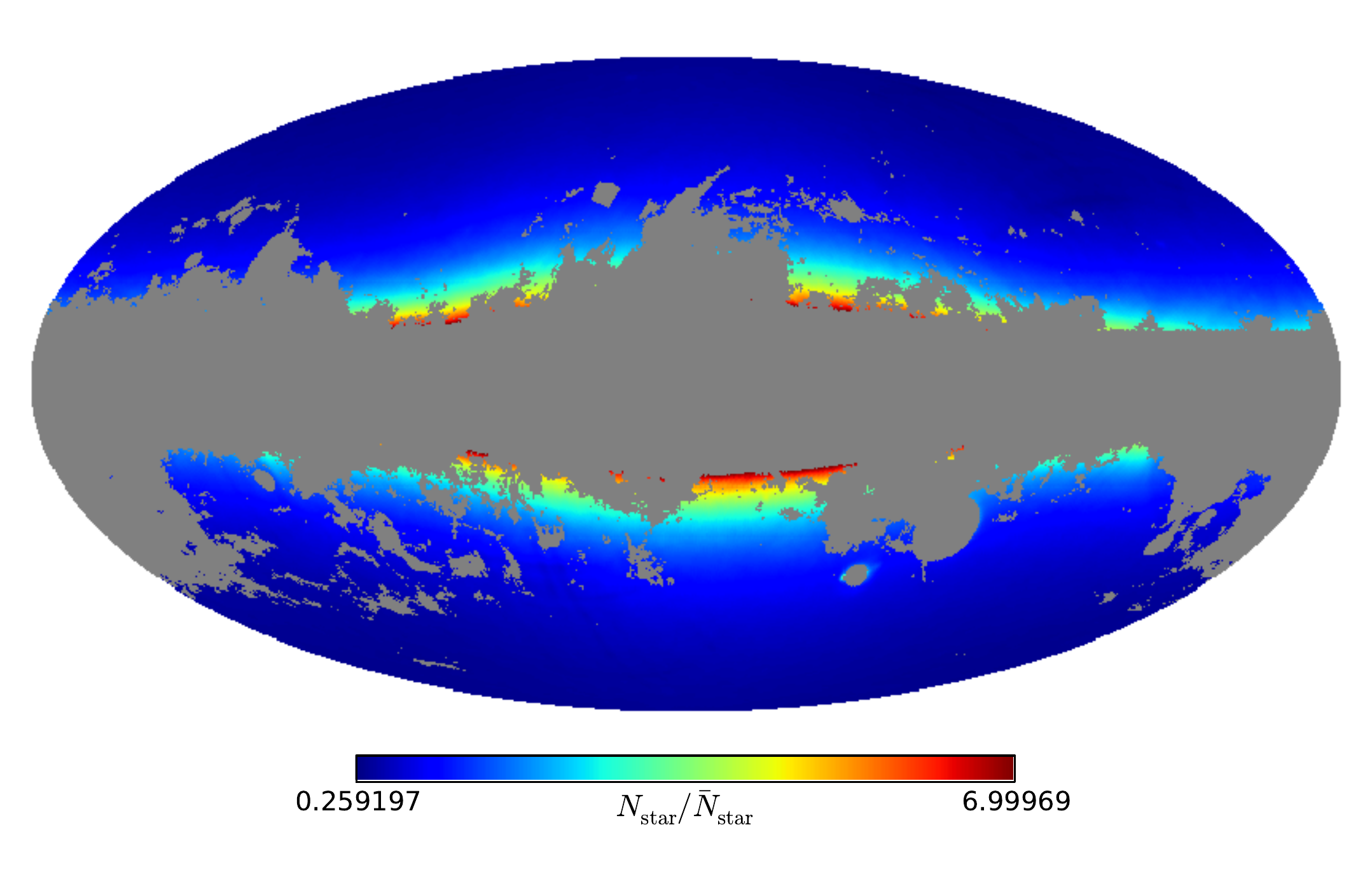}
  \caption{Stellar density template, derived from Gaia DR2 sources selected as described in Sec.~    \ref{sec:gaia}, used to estimate contamination and obscuration in WSC projected galaxy distribution.}
  \label{fig:star-template}
\end{figure}

A few unusually bright pixels -- with excessive object counts due to the presence 
of globular clusters and nearby dwarf galaxies -- can be seen in Fig.~\ref{fig:gaia-map}. To remove these count peaks, we computed, for all the pixels, the median of the number counts 
in neighboring pixels; if the central pixel value was more than twice the median, we replaced 
the former by the latter. Then, to attenuate the Poisson noise (visible in Fig.~\ref{fig:wscXsdss-star-b}), we applied Gaussian smoothing with standard deviation 
$\sigma_{\mr{G}}=34.4$ arcmin in order to retain all the structure in our template up to 
the scales of interest. Finally, we applied the same mask as in the data (see Sec.~\ref{sec:wsc}) 
and normalized the template by its mean value (the template's absolute scale does not affect the 
data treatment). The final result is shown in Fig.~\ref{fig:star-template}.

\begin{figure}
  \center
  \includegraphics[width=1.0\textwidth]{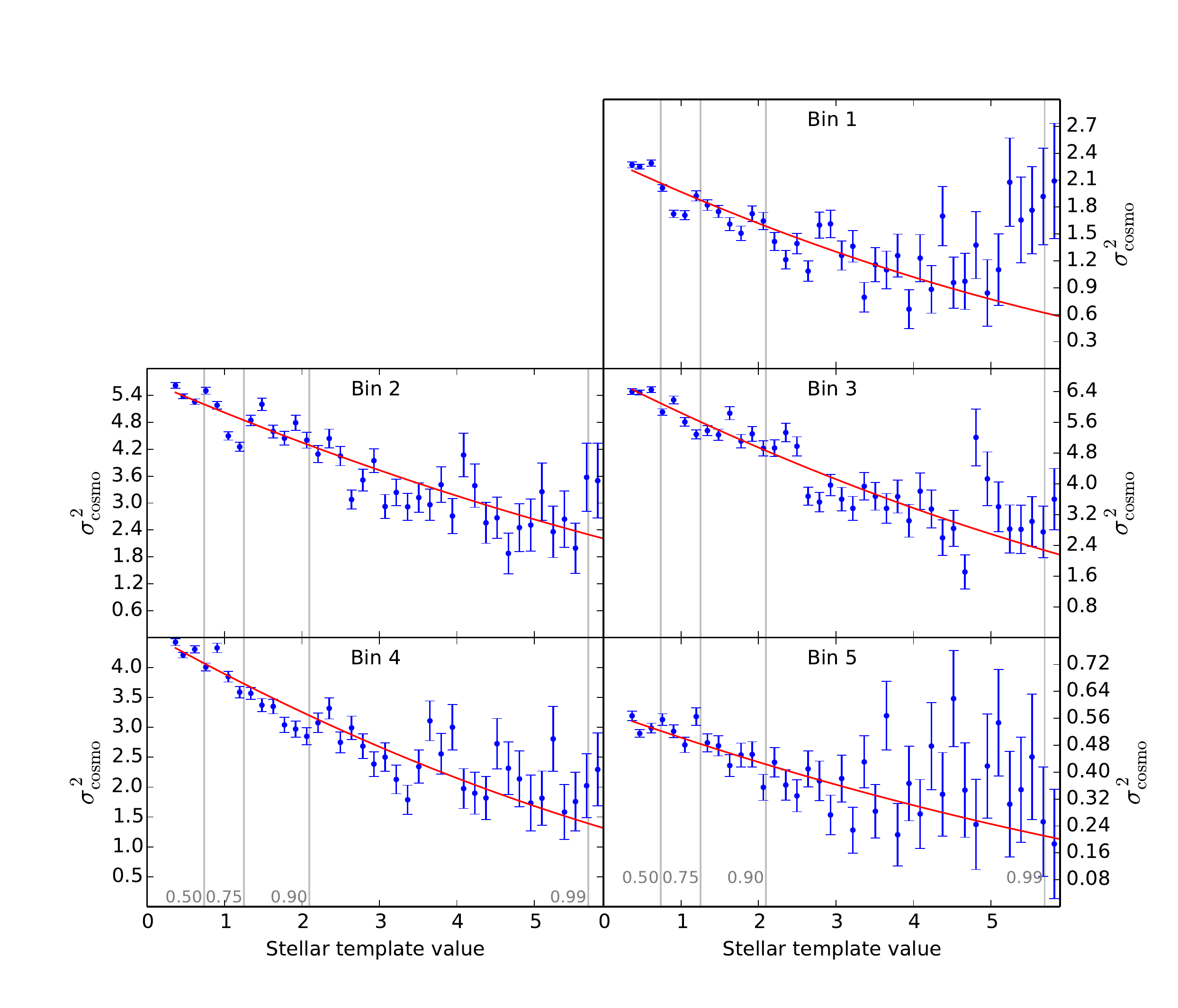}
  \caption{Variance attributed to cosmological fluctuations $\sigma_{\mr{cosmo}}^2$ as a function 
    of the stellar density template value $S(s)$ (Fig.~\ref{fig:star-template}), for each redshift bin. 
    The blue data points show the values measured from data, and the red curve is the obscuration 
    model fit (Eq.~\ref{eq:iso-model}). The vertical gray lines with labels show the survey area 
    fraction with a smaller $S(s$).}
  \label{fig:wsc-dev-gaia}
\end{figure}

Using the template from Fig.~\ref{fig:star-template}, we estimated $\alpha_i$ for all redshift bins 
with Eq.~(\ref{eq:iso-model}). Fig.~\ref{fig:wsc-dev-gaia} shows the data variance in template isocontours 
as a function of the template's value and our fit. Again, our model describes reasonably well the obscuration 
effect, apart from the first bin. 

We next removed the contamination using a slightly modified version of the 
approach proposed in \cite{Rybicki92,Elsner17}, i.e.\ we computed the correlation of the (weighted) data 
with the contamination template. The weighted number counts are defined as:

\begin{equation}
n_{\mr{w}}^i(\bm{\theta}) \equiv \frac{n_{\mr{obs}}^i(\bm{\theta})}{1-\hat{\alpha}_iS(\bm{\theta})}
\simeq W(\bm{\theta}) \left[ n_{\mr{g}}^i(\bm{\theta}) + \beta_i \frac{S(\bm{\theta})}{1-\hat{\alpha}_iS(\bm{\theta})} + 
\frac{\epsilon_{\mr{g},i}(\bm{\theta}) + \epsilon_{\mr{s},i}(\bm{\theta})}{1-\hat{\alpha}_iS(\bm{\theta})}\right].
\end{equation}
In the equation above, $\hat{\alpha}_i$ is our estimate for the obscuration parameter, and the first term inside the 
square brackets shows we assumed $\hat{\alpha}_i=\alpha_i$. Deviations from this assumption were studied in \cite{Shafer15} and 
they mostly lead to a small re-scaling of the angular power spectrum; we also verified this effect in our simulations 
(see Sec.~\ref{sec:validation}). We call the term inside the brackets that multiplies $\beta_i$ the \textit{weighted 
template}.

Now we can compute the average weighted density $\bar{n}_{\mr{w}}^i$ in the redshift bin $i$ over the survey 
footprint $W(\bm{\theta})$ and do the same for the weighted template:

\begin{equation}
\bar{S}^i_{\mr{w}} \equiv \frac{1}{A} \int \frac{S(\bm{\theta})}{1-\hat{\alpha}_iS(\bm{\theta})} \mr{d^2}\theta,
\end{equation}
where $A$ is the unmasked area of the survey. Assuming that the Poisson noise and the true galaxy overdensity 
$\delta n_{\mr{g}}^i(\bm{\theta}) \equiv n_{\mr{g}}^i(\bm{\theta}) - \bar{n}_{\mr{g},i}$ do 
not correlate with our weighted template difference $\delta S_{\mr{w}}^i(\bm{\theta}) \equiv S_{\mr{w}}^i(\bm{\theta}) - \bar{S}_{\mr{w}}^i$, 
$\beta_i$ can be estimated by:

\begin{equation}
\beta_i \simeq \frac{\int \delta n_{\mr{w}}^i(\bm{\theta}) \delta S_{\mr{w}}^i(\bm{\theta}) \mr{d^2}\theta}
     {\int [\delta S_{\mr{w}}^i(\bm{\theta})]^2 \mr{d^2}\theta},
\label{eq:compute-beta}
\end{equation}
where $\delta n_{\mr{w}}^i(\bm{\theta}) \equiv n_{\mr{w}}^i(\bm{\theta}) - \bar{n}_{\mr{w}}^i$. 
In practice, we may expect a small random correlation between $\delta n_{\mr{g}}^i(\bm{\theta})$ and
$\delta S_{\mr{w}}^i(\bm{\theta})$, which was investigated and corrected for in \cite{Elsner17}. 
We nevertheless ignore this bias as it does not affect our analysis significantly (see Sec.~\ref{sec:validation}).
Lastly, the true mean galaxy density $\bar{n}_{\mr{g},i}$ is estimated from the data as:
\begin{equation}
\hat{n}_{\mr{g},i} = \frac{\int [n_{\mr{obs}}^i(\bm{\theta}) - \hat{\beta}_i S(\bm{\theta})] \mr{d^2}\theta}
{\int W(\bm{\theta}) [1-\hat{\alpha}_iS(\bm{\theta})]\mr{d^2}\theta}.
\label{eq:compute-ng}
\end{equation}

The estimated parameters $\alpha_i$, $\beta_i$ and $\bar{n}_{\mr{g},i}$, are presented in 
Table \ref{tab:map-pars}, where we see that both obscuration and 
contamination are clearly present in all bins. Once more, the first bin shows 
higher values of systematics, and the estimated error bars are much larger as well, specially for the 
obscuration parameter. All these difficulties presented by the first bin indicate that its systematics are not as successfully modeled 
as for the other bins. On top of that, this redshift range is well covered by data with better constrained photo-$z$s, larger sky coverage, and apparently less affected by systematics -- 
namely the 2MPZ catalog \cite{Bilicki14,Balaguera17}. Thus, we decided to 
ignore the first redshift bin in the subsequent analysis.

\begin{table}
\begin{center}
\begin{tabular}{cccccc}
\hline
Bin & Photo-$z$ range & Mean spec-$z$ & $\alpha$    & $\beta$    & $\bar{n}_{\mr{g}}$ \\
\hline
1   & $0.10<z<0.15$   & 0.1280        & 0.0902(156) & 0.421(39)  & 3.223(48)        \\ 
2   & $0.15<z<0.20$   & 0.1753        & 0.0681(75)  & 0.259(28)  & 5.499(41)        \\
3   & $0.20<z<0.25$   & 0.2248        & 0.0735(56)  & 0.136(27)  & 6.601(33)        \\
4   & $0.25<z<0.30$   & 0.2719        & 0.0818(50)  & 0.095(25)  & 5.826(23)        \\
5   & $0.30<z<0.35$   & 0.3180        & 0.0674(80)  & 0.051(13)  & 1.748(08)        \\
\hline
\end{tabular}
\caption{Redshift bin number, photo-$z$ range and estimated values for the mean spectroscopic redshift and  obscuration $\alpha$, contamination $\beta$ and true mean galaxy density $\bar{n}_{\mr{g}}$ 
parameters. The uncertainties were estimated from mock realizations described in Sec.~\ref{sec:sims}.}
\label{tab:map-pars}
\end{center}
\end{table}

\begin{figure}
  \center
  \includegraphics[width=1.0\textwidth]{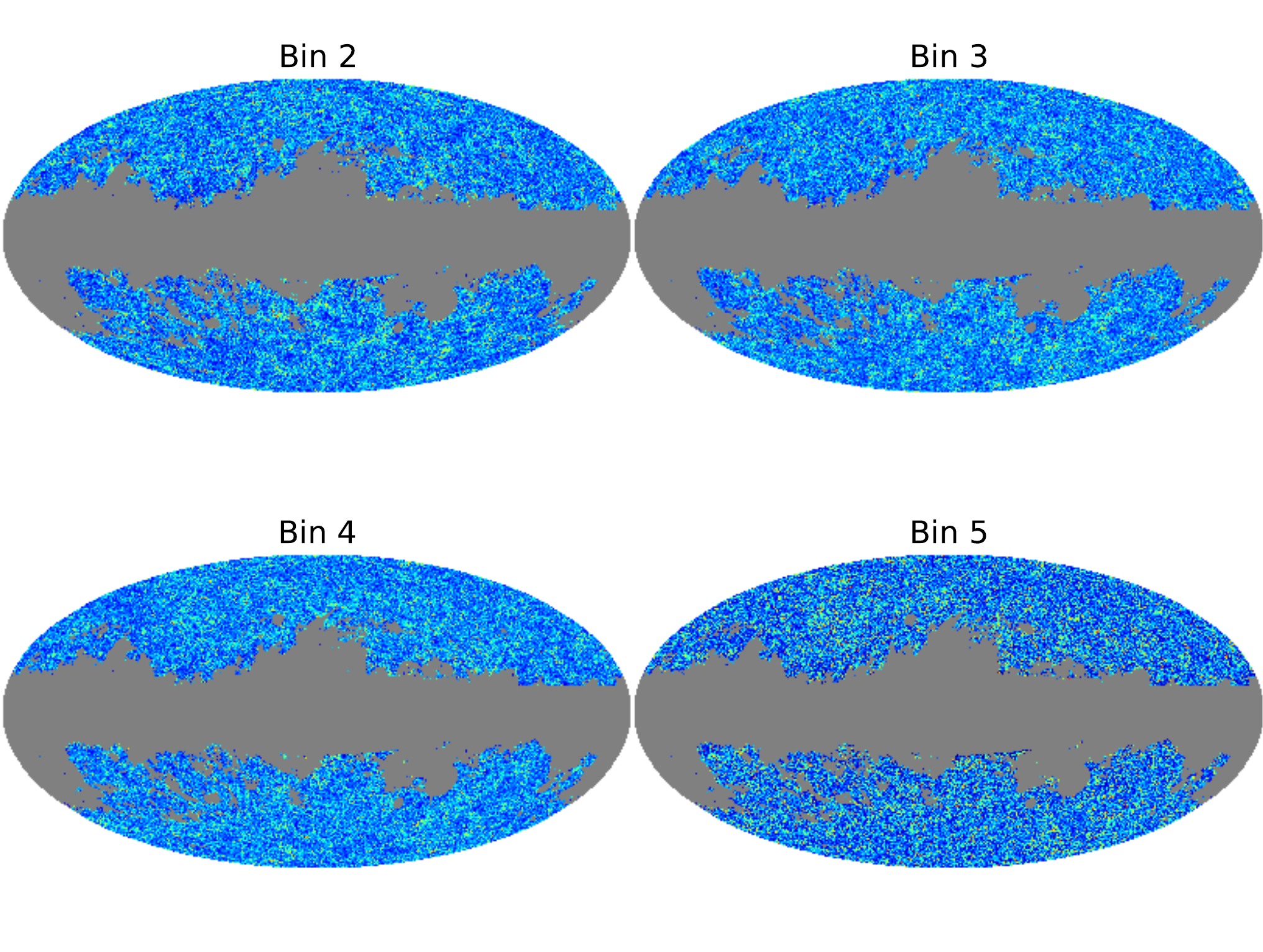}
  \includegraphics[width=1.0\textwidth]{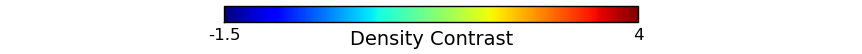}
  \caption{WSC density contrast $\delta_{\mr{g}}(\bm{\theta})$ maps for photo-$z$ 
    bins 2--5 ($0.15<z<0.35$) after correcting for stellar obscuration and contamination. 
    Due to the contamination template subtraction and Poisson fluctuations in the number counts maps, 
    some pixels end up with $\delta_{\mr{g}}(\bm{\theta})<-1$. 
  }
  \label{fig:delta-maps}
\end{figure}

Using the estimated $\alpha_i$, $\beta_i$ and $\bar{n}_{\mr{g},i}$, we can solve Eq.~(\ref{eq:sources-model}) for 
$n_{\mr{g}}^i(\bm{\theta})$ inside the survey footprint and estimate the true galaxy density contrast 
$\delta_{\mr{g}}^i(\bm{\theta})\equiv n_{\mr{g}}^i(\bm{\theta})/\bar{n}_{\mr{g},i}-1$:
 
\begin{equation}
\hat{\delta}^i_{\mr{g}}(\bm{\theta}) = W(\bm{\theta}) \left\{\frac{n^i_{\mr{obs}}(\bm{\theta}) - \hat{\beta}_i S(\bm{\theta})}
{\hat{n}_{\mr{g},i}[1-\hat{\alpha}_iS(\bm{\theta})]} - 1\right\}.
\label{eq:delta-estimate}
\end{equation}
These are shown in Fig.~\ref{fig:delta-maps}. From Eq.~(\ref{eq:delta-estimate}) we see that, even if our 
template $S(\bm{\theta})$ and our parameter estimates all perfectly match the data, the first term inside 
the curly brackets might be negative due to Poisson fluctuations in $n^i_{\mr{obs}}(\bm{\theta})$ [for instance, 
$n^i_{\mr{obs}}(\bm{\theta})$ might be zero in a given pixel].

\section{Methodology for cosmological analysis}
\label{sec:methods}

\subsection{Angular power spectrum estimation}
\label{sec:cl}

We measured the auto and cross angular power spectra $C_\ell^{ij}$ of the WSC density contrast maps using the \texttt{NaMaster}\footnote{\url{http://github.com/LSSTDESC/NaMaster}} code \cite{Alonso18}, 
which is based on a pseudo-$C_\ell$ estimator proposed in \cite{Peebles73}, and includes further 
improvements \cite{Hivon02,Elsner17}. \texttt{NaMaster} first computes the so-called pseudo power spectra $D^{ij}_\ell$, 
which measure the variance of the coefficients $a^{i}_{\ell m}$ used to expand the \emph{masked} density contrast maps 
$\delta^i_{\mr{g}}(\bm{\theta})$ in spherical harmonics $Y_{\ell m}(\bm{\theta})$: 

\begin{equation}
D^{ij}_{\ell} \equiv \frac{1}{2\ell+1}\sum_{m=-\ell}^{\ell}a^i_{\ell m} a^{j*}_{\ell m}.
\end{equation}
The coefficients $a_{\ell m}$ represent the spherical harmonic transform of the product of the mask $W(\bm{\theta})$ and the estimate of the underlying galaxy number density contrast, according to Eq.~(\ref{eq:delta-estimate}). Therefore, when transformed to harmonic space, this product translates into a convolution of the underlying (i.e., full sky) power spectrum $F^{ij}_{\ell'}$ and a mixing matrix $R_{\ell\ell'}$ \cite{Hivon02}:
 
\begin{figure}
  \center
  \includegraphics[width=1.00\textwidth]{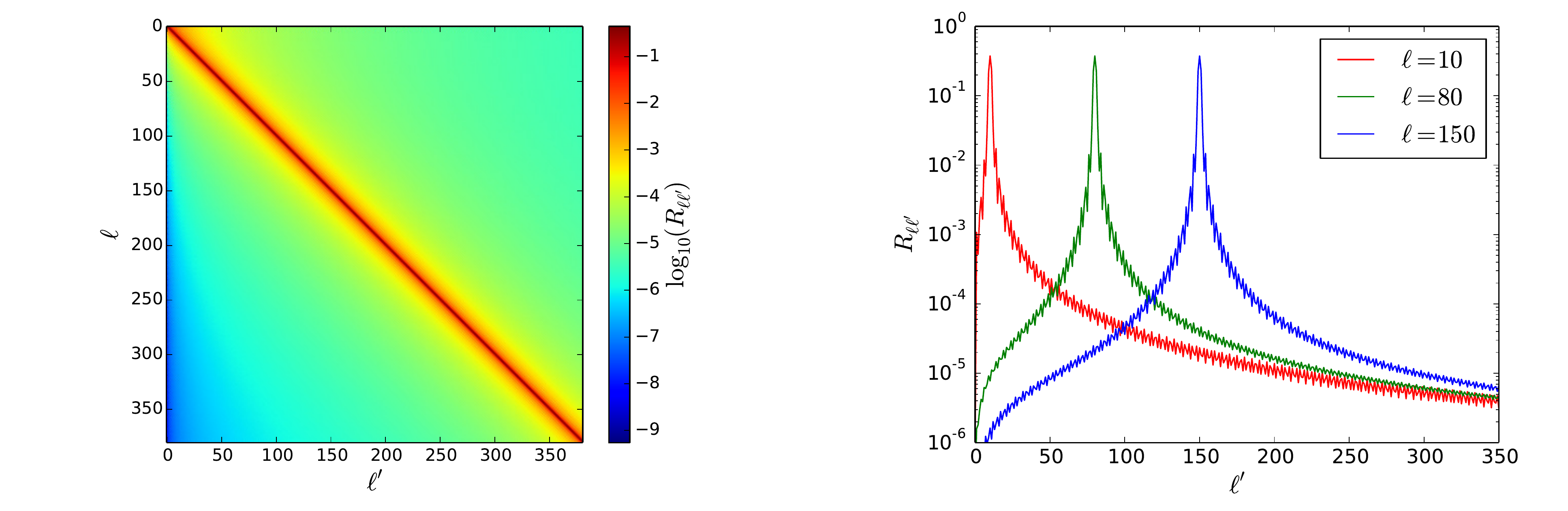}
  \caption{\emph{Left panel:} a color plot of the logarithm of the mixing matrix $R_{\ell\ell'}$ computed for our 
WSC mask, up to $\ell=380$. \emph{Right panel:} plot of $R_{\ell\ell'}$ for $\ell$ fixed at 10 (red), 80 (green) 
and 150 (blue).}
  \label{fig:rll-plot}
\end{figure} 
 
\begin{equation}
D^{ij}_{\ell} = \sum_{\ell'}R_{\ell\ell'}F^{ij}_{\ell'}.
\label{eq:mixing-matrix}
\end{equation}
The mixing matrix -- determined from the power spectrum of the mask (see e.g. equation 6 of \cite{Balaguera17}) -- 
is shown in Fig.~\ref{fig:rll-plot}, where we present on the left hand side a 2D color map of the full $R_{\ell\ell'}$, 
while on the left panel we present particular examples of the $R_{\ell\ell'}$ for three values of $\ell'$. 
Solving for $F^{ij}_{\ell'}$ by inverting the mixing matrix, our final estimate of the true, full-sky angular 
power spectra $C^{ij}_\ell$ is given by:

\begin{equation}
C^{ij}_{\ell} = \frac{F^{ij}_{\ell} - N^{ij}}{w_\ell^2},
\end{equation}
where the denominator removes the effect of the pixelization ($w_\ell$ is known as the pixel window function, 
computed with the Healpix package for our mask resolution) and $N^{ij}$ is the shot-noise contribution, given by:
\begin{equation}
N^{ij} = \frac{\delta_{\mr{K}}^{ij}}{\bar{n}_{\mr{g},i}^2} 
\int \frac{n_{\mr{obs}}^i(\bm{\theta})}{[1-\hat{\alpha}_iS(\bm{\theta})]^2} \mr{d^2}\theta\,
\label{eq:shotnoise}
\end{equation}
with $\delta_{\mr{K}}^{ij}$  the Kronecker delta. Table \ref{tab:linear-l} shows that thanks to the large mean 
number density of galaxies in most redshift bins, the shot-noise term does not dominate on the scales we are interested in. 
There are two aspects of Eq.~(\ref{eq:shotnoise}) worth pointing out. First, that these are the full number counts $n_{\mr{obs}}^i(\bm{\theta})$ (including contamination by stars)
that contribute to the shot-noise. Second, since we weight the data by a factor $[1-\hat{\alpha}_i S(\bm{\theta})]$, the 
noise is also amplified. 

We will present our measurements of the power spectrum in the form of averages in bins of width $\Delta\ell=6$, and we 
consider angular scales from $\ell_0=3$.

\subsection{Estimation of cosmological parameters}

In order to set constraints on cosmological parameters from the measured angular power spectra, we used the  
Markov Chain Monte Carlo (MCMC) code \texttt{Montepython}\footnote{\url{http://baudren.github.io/montepython.html}} 
\cite{Audren12,Brinckmann18}.
We have adopted a flat $\Lambda$CDM cosmological model with the following free parameters: 
$h=H_0/(100\; \mr{km~s^{-1}~Mpc^{-1}})$, where $H_0$ is the Hubble constant; 
$\Omega_{\mr{b}}$, the baryon density parameter; $\Omega_{\mr{c}}$, the cold dark matter density parameter; 
$\ln(10^{10}A_{\mr{s}})$, where $A_{\mr{s}}$ is the amplitude of the primordial power spectrum; $n_{\mr{s}}$, 
the spectral index of the primordial power spectrum; the galaxy biases $b_i$ in each of the four redshift 
bins $i$; and the widths $\sigma_{\mr{z},i}$ of the $i$th bin's spec-$z$ selection function 
as nuisance parameters;
in total 13 parameters. We also computed the amplitude of matter density fluctuations $\sigma_8$ as a derived parameter. 
As priors we used independent Gaussian distributions for parameters 
poorly constrained by our data: $h$, $\ln(10^{10}A_{\mr{s}})$ and $n_{\mr{s}}$, with means and standard deviations 
according to Planck \cite{Planck16}. We point out that our narrow prior on $\ln(10^{10}A_{\mr{s}})$ is translated 
into prior constraints on $\sigma_8$ as well.

The theoretical $C^{ij}_\ell$s used in the MCMC were computed by 
\texttt{CLASS}\footnote{\url{http://class-code.net}} \cite{Blas11,Dio13}, including contributions from 
redshift space distortions and non-linear structure growth \citep{Smith03,Takahashi12}. Together with the Planck priors, 
we applied the extra restriction $\Omega_{\mr{b}}>0.0065$, to avoid a crash in 
\texttt{CLASS} caused by reionization happening at too late times. 
To describe the spec-$z$ distribution of galaxies in each photo-$z$ bin, we used their Gaussian approximation; 
the impact of assuming a generalized Lorentzian distribution for the spec-$z$s is discussed in Sec.~\ref{sec:lorentzian-photo-z}.

To reduce the number of possible sources of systematic deviations between measured and true parameters, 
we restricted our analysis to linear scales. More specifically, for each $C^{ij}_\ell$ we only kept the multipoles 
for which the contribution from the non-linear part of the matter power spectrum was smaller than 10\%.
Table \ref{tab:linear-l} shows this maximum $\ell$ for each angular power spectrum.

\begin{table}
\begin{center}
\begin{tabular}{cccc}
\hline
Bin $i$ & Bin $j$ & $\ell_{\mr{max}}$ & $\ell_{\mr{shot}}$\\
\hline
2 & 2 & 70 & 362 \\
2 & 3 & 80 & -- \\
2 & 4 & 89 & -- \\
2 & 5 & 103 & -- \\
3 & 3 & 93 & 367 \\
3 & 4 & 104 & --\\
3 & 5 & 117 & -- \\
4 & 4 & 117 & 316 \\
4 & 5 & 129 & -- \\
5 & 5 & 140 & 180 \\
\hline
\end{tabular}
\caption{Maximum multipole (i.e. linear scales) $\ell_{\mr{max}}$ included in our MCMC estimation of the posterior, and the multipole at which the shot noise starts to dominate, $\ell_{\mr{shot}}$, for each $C^{ij}_\ell$.}
\label{tab:linear-l}
\end{center}
\end{table}

\begin{figure}
  \center
  \includegraphics[width=1.0\textwidth]{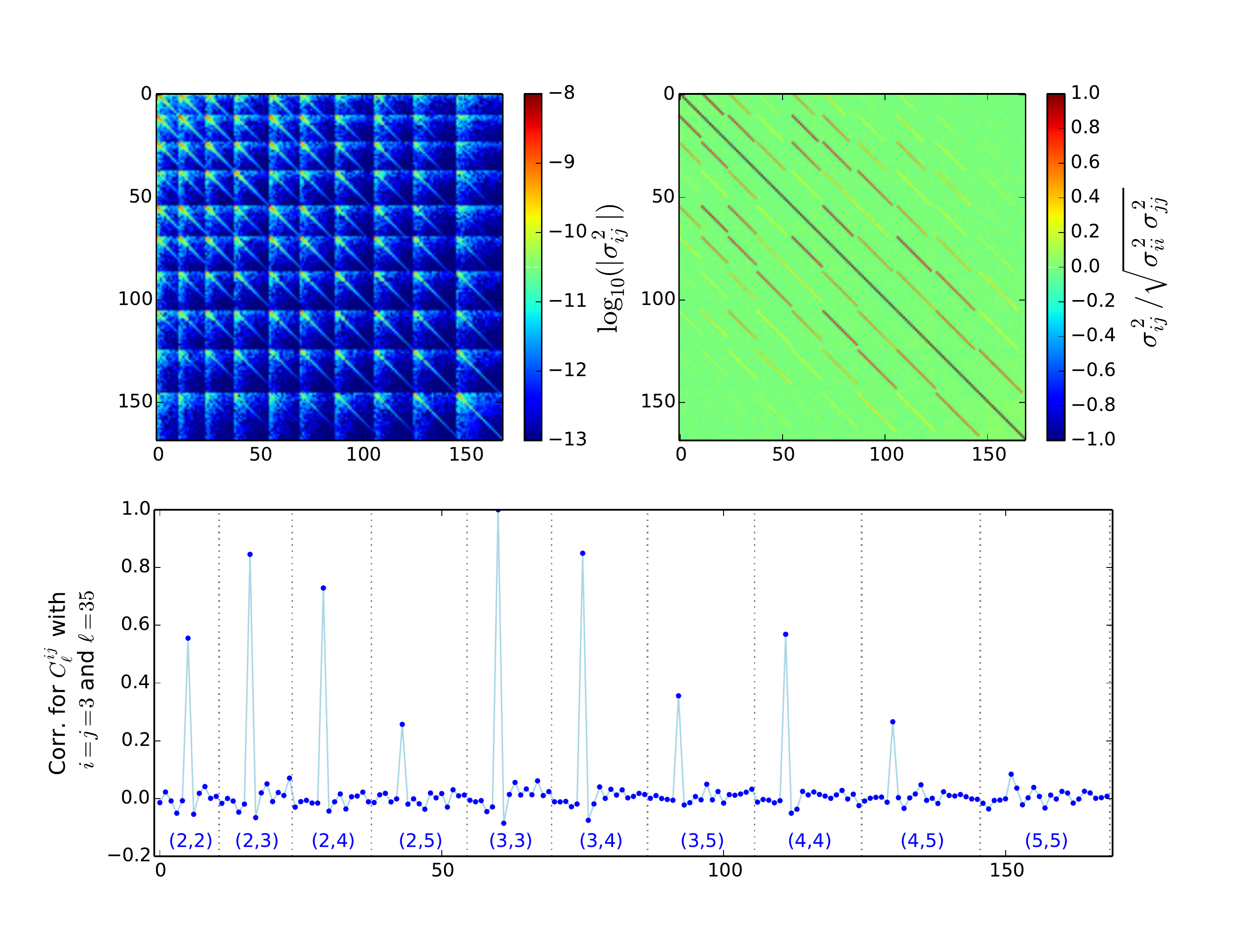}
  \caption{\emph{Top panels:} logarithm of the absolute value of our data vector's covariance matrix (left) 
      and the data vector's correlation matrix (right). The data vector is composed of 
      10 binned auto- and cross-$C_\ell$s limited to linear scales, concatenated in the following order:   
      $C_\ell^{22}, C_\ell^{23},\ldots, C_\ell^{25}, C_\ell^{33}, C_\ell^{34}, \ldots, C_\ell^{55}$. 
      The left-hand plot was clipped at $-13$ to improve visualization (its largest values are unclipped). The covariances 
      between $C_\ell^{ij}$ and $C_\ell^{kq}$ appear as sub-matrices. \emph{Bottom panel:} 60th row of the 
      correlation matrix, corresponding to $C^{ij}_\ell$ with $i=j=3$ and $\ell=35$. The sectors corresponding 
      to the pair of redshift bins $i$ and $j$ are separated by vertical dotted lines and labeled by $(i,j)$. 
      We note that different  (binned) $\ell$s are practically uncorrelated, and that the power spectra 
      present stronger covariances when they involve nearby bins.
  }
  \label{fig:cov-matrix}
\end{figure}

The binned $C^{ij}_\ell$s were combined into a single data vector with 169 entries. 
We assumed a Gaussian likelihood, with the covariance (see Fig.~\ref{fig:cov-matrix}) extracted from 2500 
lognormal mock catalogs of the WSC, described in Sec.~\ref{sec:sims}. The $C^{ij}_\ell$s are measured from these mocks, 
following the same procedure as used for the data (described 
in Secs.~\ref{sec:cleaning} and \ref{sec:cl}). The inverse of the covariance matrix estimated from the simulations 
has been corrected for the finite sample used to estimate it, following \cite{Hartlap07}.

The sampling of the posterior was performed with 10 Markov chains of 100,000 entries, 
each one starting from a different point in the parameter space. The convergence of the chains was verified 
using the Gelman-Rubin diagnostic $R$, and all parameters presented $R<1.15$. 

\section{Validation of the analysis}
\label{sec:validation}

To make sure that all the steps of our analysis work properly, we used the Full-sky Lognormal Astro-fields 
Simulation Kit (FLASK\footnote{\url{http://www.astro.iag.usp.br/~flask}}) \cite{Xavier16} 
to create hundreds of mock WSC catalogs including the obscuration and contamination effects 
described in Sec.~\ref{sec:character}. These mocks are described in detail in Sec.~\ref{sec:sims}.
The results obtained with our methodology are shown in Sec.~\ref{sec:par-recovery}.

\subsection{Mock WSC maps}
\label{sec:sims}

To create lognormal realizations that reproduce the basic statistical properties of the data, FLASK 
may use as input: a set of auto- and cross-$C_\ell$s for all the redshift slices being simulated; the 
mean density of sources $\bar{n}_{\mr{g},i}$; and the angular completeness $f(\bm{\theta})$, which 
includes the mask, obscuration and/or angular variations in density. FLASK output is a set of 
Healpix maps that include Poisson noise and that reproduce all one- and two-point statistics as 
required, including cross-correlations between redshift bins. It can also create realizations of 
stellar densities by assuming $C_\ell^{\mr{star}}=0$, setting $f(\bm{\theta})$ to our template 
and Poisson sampling from it. 

\begin{figure}
  \center
  \includegraphics[width=0.49\textwidth]{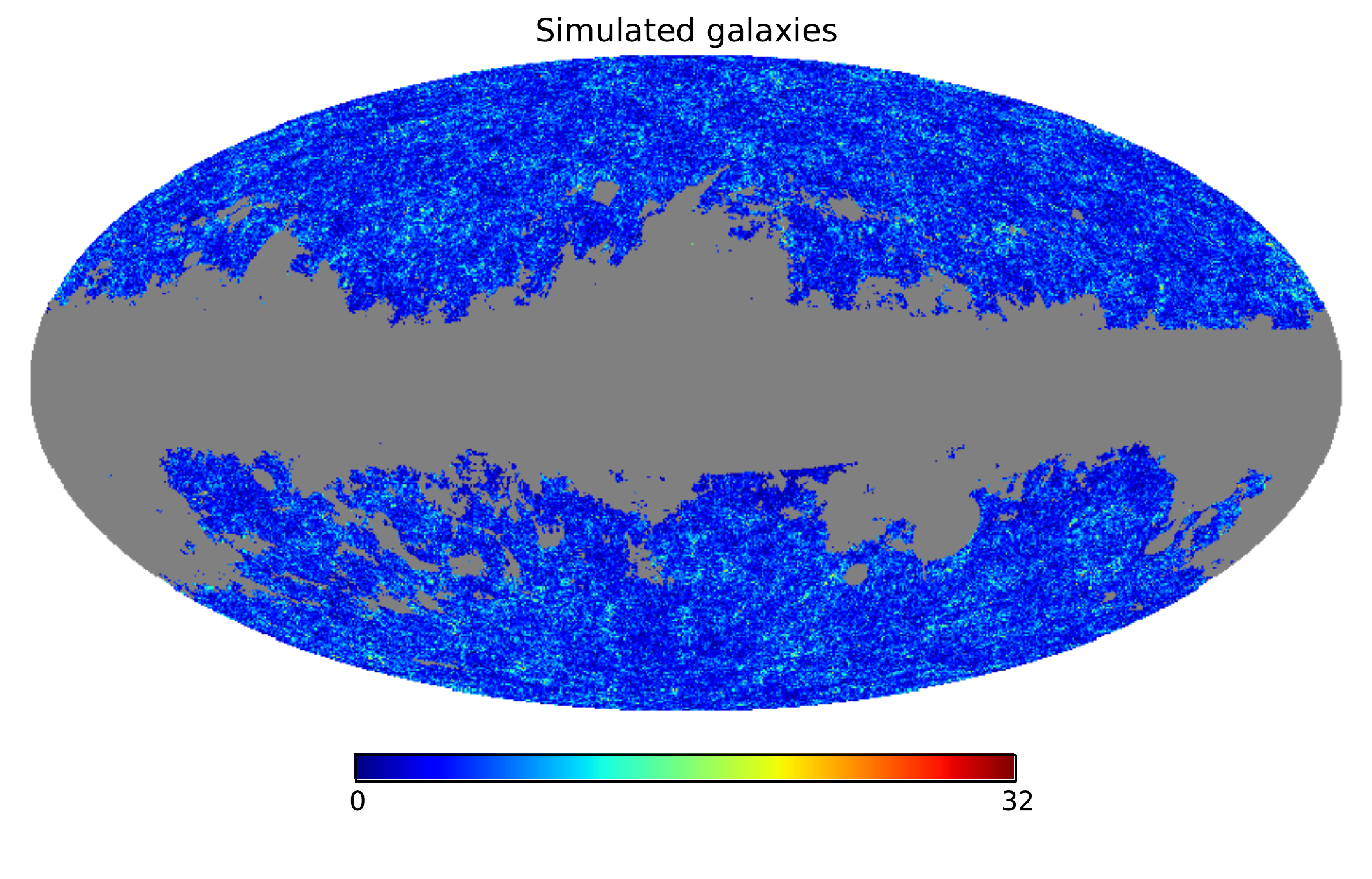}
  \includegraphics[width=0.49\textwidth]{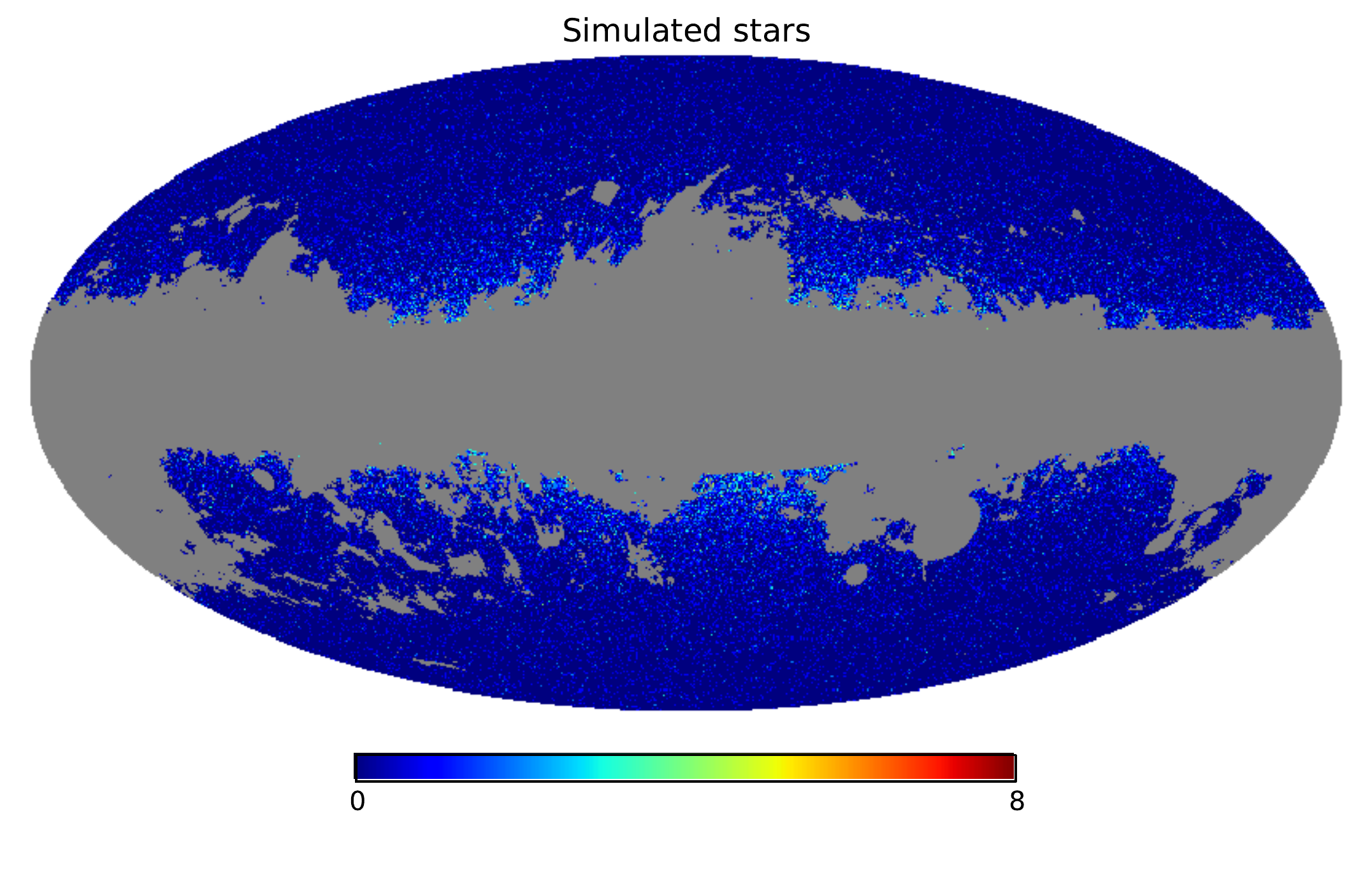}
  \includegraphics[width=0.49\textwidth]{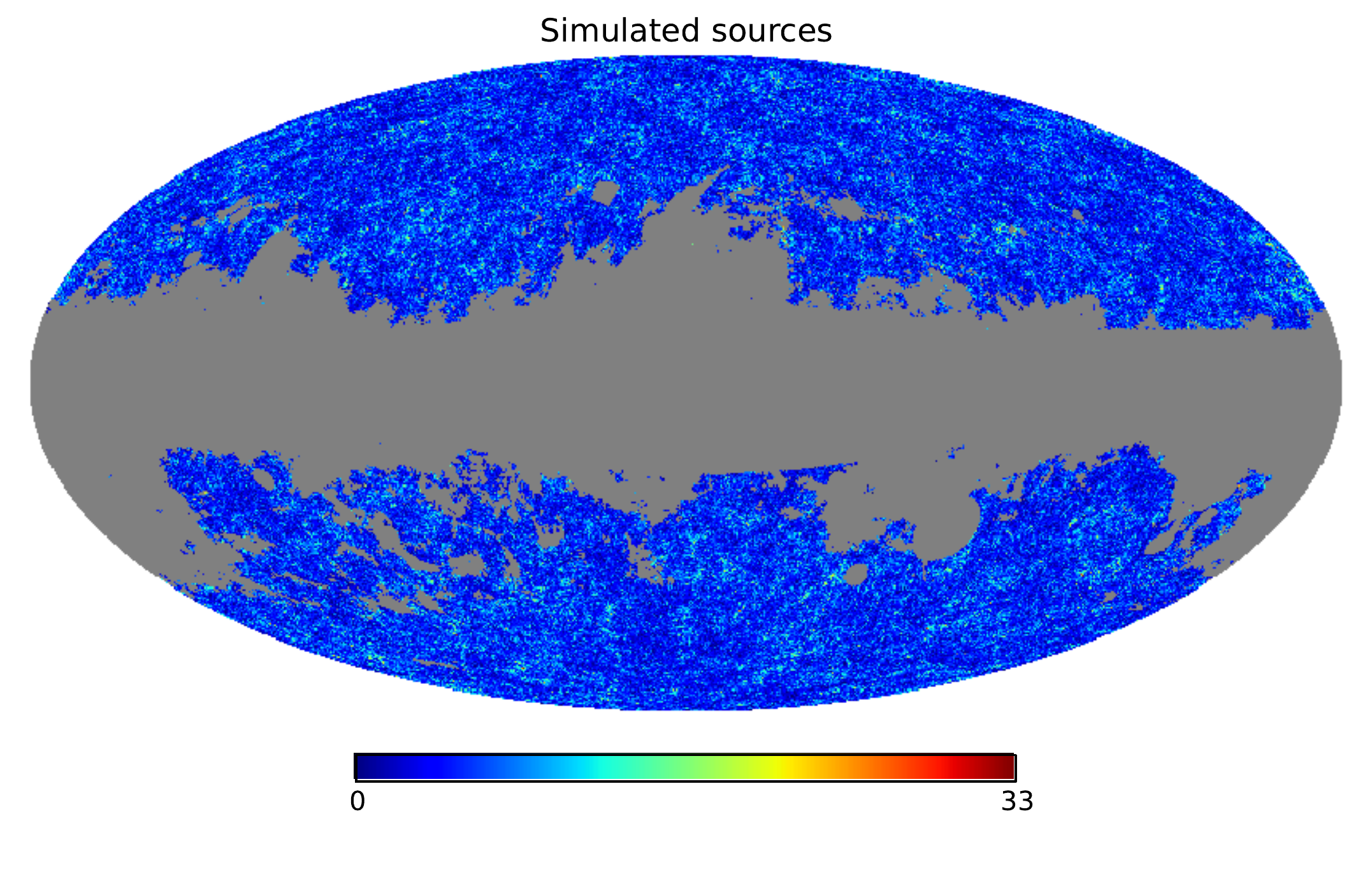}
  \caption{\emph{Top left panel:} a lognormal realization of WSC galaxies following
    $C_\ell^{ij}$, $\bar{n}_{\mr{g},2}$ and $\alpha_2$ [it includes obscuration according to the 
    stellar density template $S(\bm{\theta})$]. \emph{Top right panel:} a realization of a stellar 
    density map, made by Poisson sampling  $S(\bm{\theta})$ scaled by $\beta_2$. \emph{Bottom panel:} the combination 
    of both maps in the top panels, representing all sources detected in the photo-$z$ bin 2 
    ($0.15<z<0.20$).}
  \label{fig:sim-maps}
\end{figure}

As the expected densities (angular-position dependent), we used $\bar{n}_{\mr{g},i}[1-\alpha_iS(\bm{\theta})]$ for 
galaxies and $\beta_iS(\bm{\theta})$ 
for stars, where $S(\bm{\theta})$ is our template (Fig.~\ref{fig:star-template}) and the remaining 
parameters are set according to Table \ref{tab:map-pars}. 
As input galaxy $C_\ell$s we used smooth fits to the measured cleaned $C_\ell$s, presented later in Fig.~\ref{fig:data-cls}, 
up to $\ell_{\mr{max}}=640$. 
The maps of galaxies and stars, for each bin, 
were summed into a single map of sources, to which we applied our analysis. Fig.~\ref{fig:sim-maps}
shows an example of such a realization. 
 
\subsection{Recovery of mock properties}
In this Section we test our methodology to account for the systematic effects and the likelihood machinery on our mock catalogs.
\label{sec:par-recovery}

\subsubsection{Obscuration and contamination parameters}

The first step in our method is to estimate the obscuration $\alpha_i$ and contamination $\beta_i$ 
parameters for all bins $i$, along with the mean galaxy density $\bar{n}_{\mr{g},i}$. Table \ref{tab:recov-pars} 
shows the mean and standard deviation of these parameters, recovered from 2500 WSC mock realizations. 
It shows that our method traces quite well the true values in each of the bins, being able to detect 
obscuration and contamination at no less than 5.7$\sigma$ and 3.9$\sigma$, respectively. 
It does, however, 
produce a detectable average bias of $\sim3\%$ on $\alpha$ and $\sim7\%$ on $\beta$ (i.e. overcorrecting, 
on average, both effects). 
This bias is likely caused by our simplifying assumption that the cosmological 
variance $\sigma_{\mr{cosmo}}^2(s)$ for counts-in-pixels, computed from a pixel set, does not depend on the template's isocontour 
(Eq.~\ref{eq:iso-var}), which happens to vary both in shape and in area. As density fluctuations are 
spatially correlated, variances computed in a smaller and compact set will, in general, be smaller. 
Since, as we show next, this bias does not significantly 
affect our results, we considered the method's performance adequate despite of its presence. 

\begin{table}
\small
\begin{center}
\begin{tabular}{c|cccc|cccc|cccc}
\hline
Bin&$\alpha_0$&$\bar{\alpha}$&$\sigma_\alpha$&\% bias&$\beta_0$&$\bar{\beta}$&$\sigma_\beta$&\% bias&$n_{\mathrm{g}0}$&$\bar{n}_\mathrm{g}$&$\sigma_{n_\mathrm{g}}$&\% bias\\
\hline
1&0.0902&0.0934&0.0157&3.4&0.421&0.435&0.039&3.2&3.223&3.22&0.049&-0.1\\
2&0.0681&0.0705&0.0077&3.5&0.259&0.273&0.029&5.3&5.499&5.498&0.042&-0.0\\
3&0.0735&0.0755&0.0056&2.6&0.136&0.15&0.028&9.0&6.601&6.601&0.033&-0.0\\
4&0.0818&0.0836&0.005&2.2&0.095&0.106&0.024&10.5&5.826&5.826&0.023&0.0\\
5&0.0674&0.0703&0.0081&4.0&0.051&0.056&0.013&8.8&1.748&1.748&0.008&0.0\\
\hline
\end{tabular}
\end{center}
\caption{Comparison of true obscuration, contamination and mean galaxy density parameters and those recovered from 2500 simulations. 
The columns are: the redshift bin and the true value $x_0$, the mean value from the simulations $\bar{x}$, the standard deviation 
of the simulations $\sigma_x$, and the fractional bias ($\bar{x}/x_0-1$) in percent for each parameter $x$.}
\label{tab:recov-pars}
\end{table}

\subsubsection{Angular power spectra}
\label{sec:cl-validation}
\begin{figure}
  \center
  \includegraphics[width=0.85\textwidth]{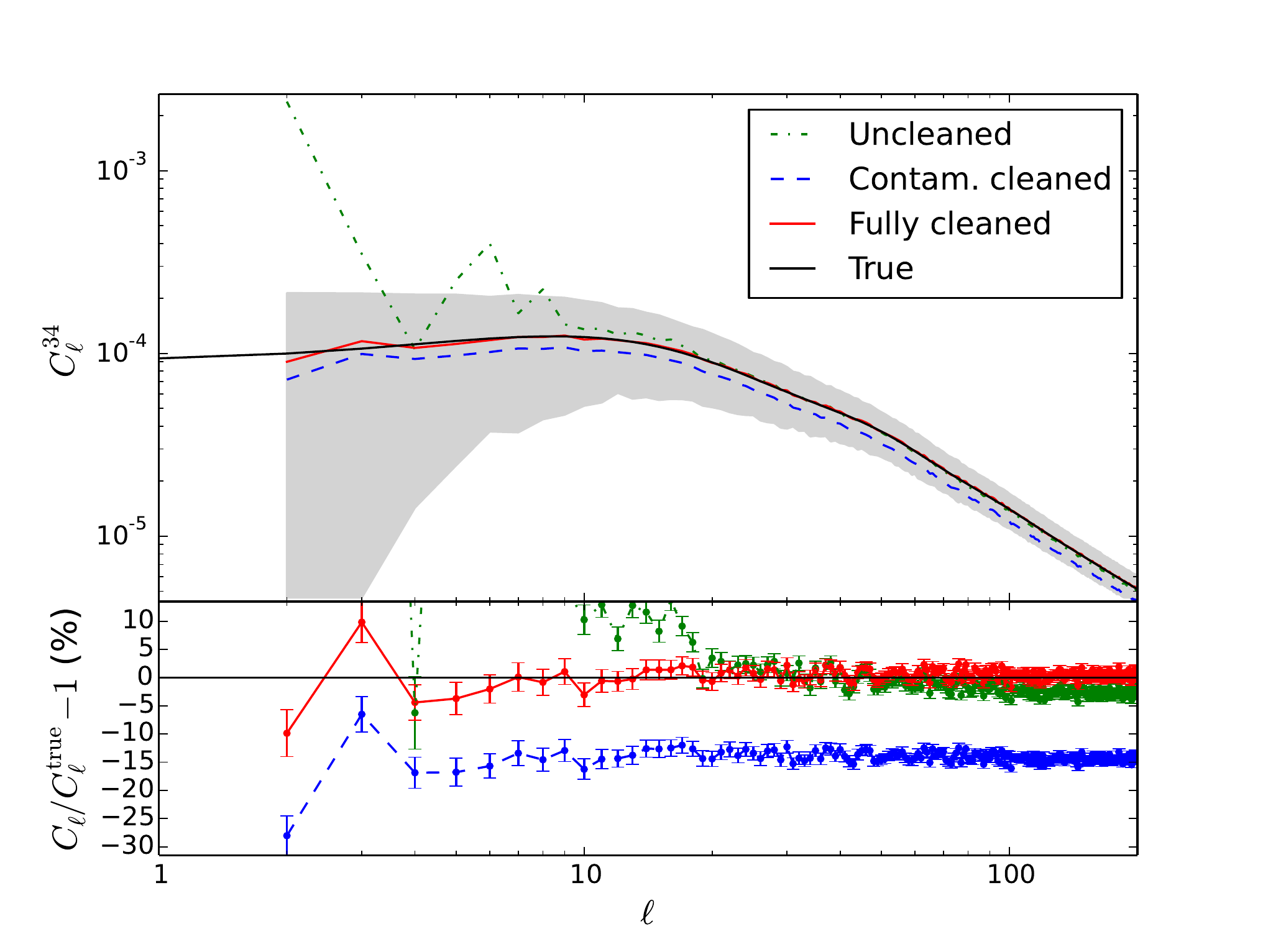}
  \caption{\emph{Top panel:} cross-$C_\ell$ from simulations for the redshift bins 3 and 4. The smooth 
    black curve represents the simulation input $C_\ell$, while the remaining 
    lines represent the average over 750 $C_\ell$s extracted from simulated maps that     
    include obscuration and contamination. The green dot-dashed line shows the results 
    of ignoring both systematic effects; the blue dashed line shows the results when only 
    contamination is corrected for; and the solid red line shows the results when both effects 
    are accounted for. The gray shaded area represents the typical uncertainty over one single $C_\ell$ 
    measurement, and is centered on the input $C_\ell$.    
    \emph{Bottom panel:} the fractional difference between the average of the 750 estimated
    $C^{34}_\ell$s and its true value, for all the cleaning strategies. The error bars show the uncertainty 
    of the mean.}
  \label{fig:sim-cls}
\end{figure}

In order to verify the impact of our cleaning method on the $C_\ell$s, we processed 750 simulations in three different 
ways\footnote{Due to the large amount of time required to measure 10 $C_\ell$s in three different ways from each 
simulation, we did not use all 2500 at this step.}: 
(\emph{i}) we ignored the obscuration and contamination present in them (equivalent to assuming 
$\alpha=\beta=0$ even though they are not); (\emph{ii}) we only corrected for contamination (equivalent to assuming 
$\alpha=0$); and (\emph{iii}) we applied the full cleaning method described in Sec.~\ref{sec:cleaning}. 
We then measured the angular power spectra using our standard procedure (Sec.~\ref{sec:cl}).
Fig.~\ref{fig:sim-cls} shows the average of 750 estimated $C_\ell$s obtained after applying the three different 
cleaning processes in comparison to the true $C_\ell$ in the case of the cross-correlation 
between redshift bins 3 and 4, as an example. The general characteristics of the results are the same for all bin pairs.
We see that, in our simulations, the typical effect of contamination is to increase the power on the largest scales 
($\ell\lesssim 20$), showing at $\ell\lesssim 10$ deviations with respect to the true spectrum larger than the 
$1\sigma$ error bars. This is clearly a systematic effect we need to correct for in the data. Similarly, we see that 
correcting only for contamination gives an estimate of the angular power spectrum which is biased low. 
This bias, also reported in \cite{Shafer15}, amounts to  $\sim 15\%$ and reaches the size of the error bars on 
small scales. After correcting for obscuration, this bias is eliminated (although there seems to be a small positive 
and constant bias of $\sim1\%$, possibly caused by an overestimated $\alpha$). This shows that when measuring 
cosmological or astrophysical parameters controlling the amplitude of the angular clustering signal (e.g. galaxy 
bias and primordial spectrum amplitude), correcting for obscuration is a key step to obtain accurate results. This 
comparison shows that, assuming our modeling of the systematic is correct, its application significantly improves 
the accuracy of the estimated power spectrum, leaving only possible residual bias which amounts to no more 
than $2\%$ of the error bars.

\subsubsection{Cosmological parameters}
\label{sec:mcmc-validation}

    Once we have quantified that our cleaning method properly corrects the measured power spectrum, we use the MCMC methodology 
to recover the input cosmological and nuisance parameters from cleaned $C_\ell$s. 
The input $C_\ell$s for the FLASK code were computed by CAMB 
sources\footnote{\url{http://camb.info/sources}} \cite{Challinor11}, including not only contributions 
from redshift space distortions and non-linearities, but also from a minimal neutrino configuration and 
all available effects (e.g. gravitational lensing and general relativistic 
corrections). Furthermore,  the computation of the input $C_\ell$s did not use the Limber approximation and adopted 
higher precision than the $C_\ell$s sampled by the MCMC. 
In this way, we verified that any potential deviations in the recovered parameters are not caused by approximation schemes 
or by the omission of small physical effects. The input $C_\ell$s were computed assuming a flat $\Lambda$CDM model and Gaussian 
spec-$z$ selection function, and the simulations included the effects of obscuration and contamination. The procedure followed 
here mimics that applied to the real data.

\begin{figure}
  \center
  \includegraphics[width=1.0\textwidth]{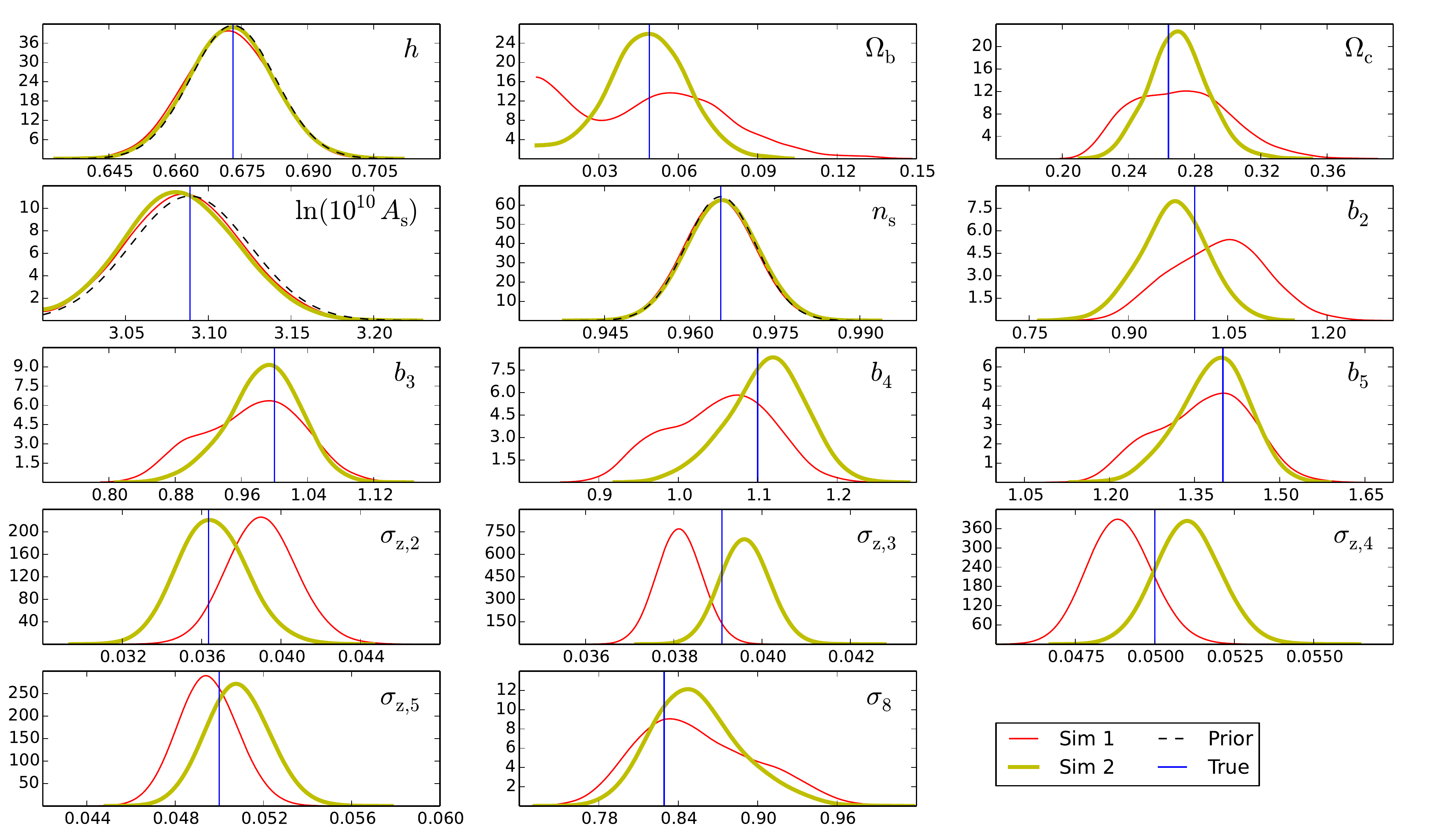}
  \caption{Performance of two example simulations in recovering input parameters.  The solid (thin red and thick ochre) 
    curves show the marginal posterior distributions for each cosmological or nuisance parameter 
    given a set of $C^{ij}_\ell$s extracted from two independent realizations of source count maps. 
    The dashed black curves show the Gaussian priors used in some parameters 
    [$h$, $\ln(10^{10}A_{\mr{s}})$ and $n_{\mr{s}}$], and the blue vertical lines depict the true values used in the 
    simulations.}
  \label{fig:sim-mcmc-marginal}
\end{figure}

Figure \ref{fig:sim-mcmc-marginal} shows that all marginal posterior distributions for all the parameters measured from two independent
simulations are compatible with their input values, thus validating our pipeline. We also point out that, while the 
covariance matrix used in our likelihood was estimated from simulations based on smooth fits to the measured  $C^{ij}_\ell$s, 
the simulated data used in this test was generated from a different set of $C^{ij}_\ell$s, thus attesting that our covariance 
matrix does not bias the results. 
The reader might notice that posterior distribution for e.g. $\sigma_{\mr{z},3}$ from one simulation is displaced from its true 
value. In this case, we point out that we have 13 independent parameters and, in fact, we should expect a few to deviate from 
their expected values by chance. The second simulation shows that this is indeed the case. Moreover, we see that the ability to 
pinpoint $\Omega_{\mr{b}}$ is affected by the specific realization of noise and cosmological fluctuations. 
 
\section{Results}
\label{sec:results}

\begin{figure}
  \center
  \includegraphics[width=1.0\textwidth]{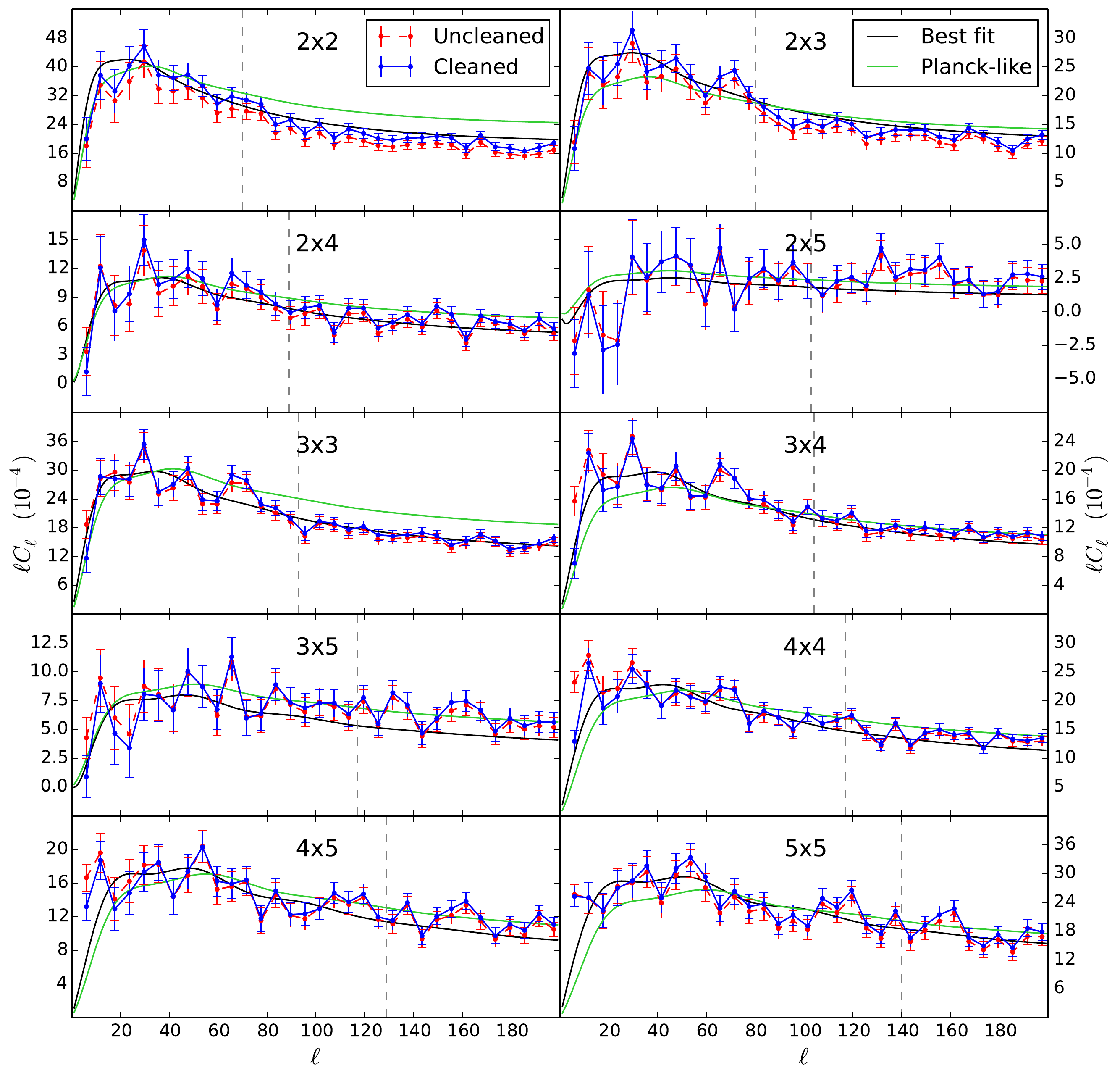}
  \caption{Binned auto and cross angular power spectra of the WSC sources. The two correlated redshift bins  
    are indicated in each plot as $i \times j$. The power spectra were multiplied by the effective $\ell$ to improve 
    readability. The red data points and dashed lines represent the $C_\ell$s extracted directly from the data, 
    without correcting for contamination and obscuration. The blue data points and lines represent the $C_\ell$s 
    taking into account such corrections. The black curves show the best fit model to the cleaned data. The green 
    curves show a $\Lambda$CDM model with Planck values \cite{Planck16} 
    and nuisance parameters set by hand, as a guiding reference. The vertical dashed gray lines mark the linear 
    limit up to which the data points were fitted.}
  \label{fig:data-cls}
\end{figure}

In Fig.~\ref{fig:data-cls} we show the binned angular power spectra measured from the WSC data, both cleaned 
and uncleaned for contamination and obscuration. We see that the effect of cleaning the WSC is 
reasonably small on most scales, indicating that galaxy catalogs containing systematics can still provide 
cosmological information. Despite the smallness of the corrections, we point out two relevant 
aspects of them: first, applying them significantly reduces the power on the largest scales ($\ell\lesssim 10$) 
in many cases, and as we showed in Sec.~\ref{sec:par-recovery}, this is caused by contamination removal; 
second, some cleaned $C_\ell$s show an almost constant-factor increase in power at all scales, caused by 
obscuration correction, as we have previously discussed (see Fig.~\ref{fig:sim-cls}).

Figure \ref{fig:data-cls} also shows the best-fit model to the cleaned data (black lines), as well as
$C_\ell$s obtained using cosmological parameters given by Planck \cite{Planck16} and nuisance parameters 
fitted ``by eye'' as a rough reference (green lines). By comparing them we see that WSC $C_\ell$s present steeper 
slopes than the fiducial cosmology on the linear scales, and that these slope differences do not seem to be 
restricted to the largest scales only ($\ell\lesssim 20$).

A comparison of the best-fit $\chi^2$ obtained under different data processing strategies 
is presented in Table \ref{tab:chi2}. In every case, the best-fit $C_\ell$s and $\chi^2$ were 
obtained for that particular dataset by fitting all 13 parameters to it.
We see that correcting the WSC count maps for obscuration and 
contamination improves the fit, but the model remains not very representative of the data. If we 
ignore multipoles $\ell<15$, the fit improves significantly and reaches a $p$-value level that 
could be considered adequate (larger than $5\%$). This may indicate that the largest scales 
are likely still contaminated even after our corrections, possibly due to a mismatch between 
the real contamination and our template. If we push the multipole cut to higher values, the agreement 
does not improve significantly, suggesting that the best-fit is not constrained by the largest 
scales after this point.   

\begin{table}
\begin{center}
\begin{tabular}{lccc}
\hline
Approach           & $\chi^2$ & d.o.f. & $p$-value\\
\hline
Uncleaned          & 268     & 156    & $5.87\times10^{-8}$ \\
Cleaned            & 237     & 156    & $2.92\times10^{-5}$ \\
$\ell\geq15$       & 163     & 136    & $0.058$ \\   
$\ell\geq21$       & 151     & 126    & $0.063$ \\
\hline
North uncleaned    & 398     & 156    & $<1\times 10^{-16}$ \\
North cleaned      & 376     & 156    & $<1\times 10^{-16}$ \\
North $\ell\geq15$ & 148     & 136    & $0.23$ \\
\hline
South uncleaned    & 248 & 156    & $3.77\times10^{-6}$ \\
South cleaned      & 185 & 156    & $0.054$ \\
South $\ell\geq15$ & 157      & 136    & $0.10$ \\ 
\hline
\end{tabular}
\end{center}
\caption{Summary statistics showing the performance of different processing strategies and best-fit models used to 
fit the angular power spectra of the WSC data. The first column names the strategy adopted. The second displays the best-fit 
$\chi^2$, computed with respect to the $C_\ell$s that best-fit that particular processing strategy. 
The third column contains the number of degrees of freedom, while the last column displays the $p$-value. 
All the approaches shown in the table that are restricted to $\ell\geq 15$ (or $\ell\geq 21$) 
were also subjected to the cleaning process.}
\label{tab:chi2}
\end{table}

\begin{figure}
  \center
  \includegraphics[width=1.0\textwidth]{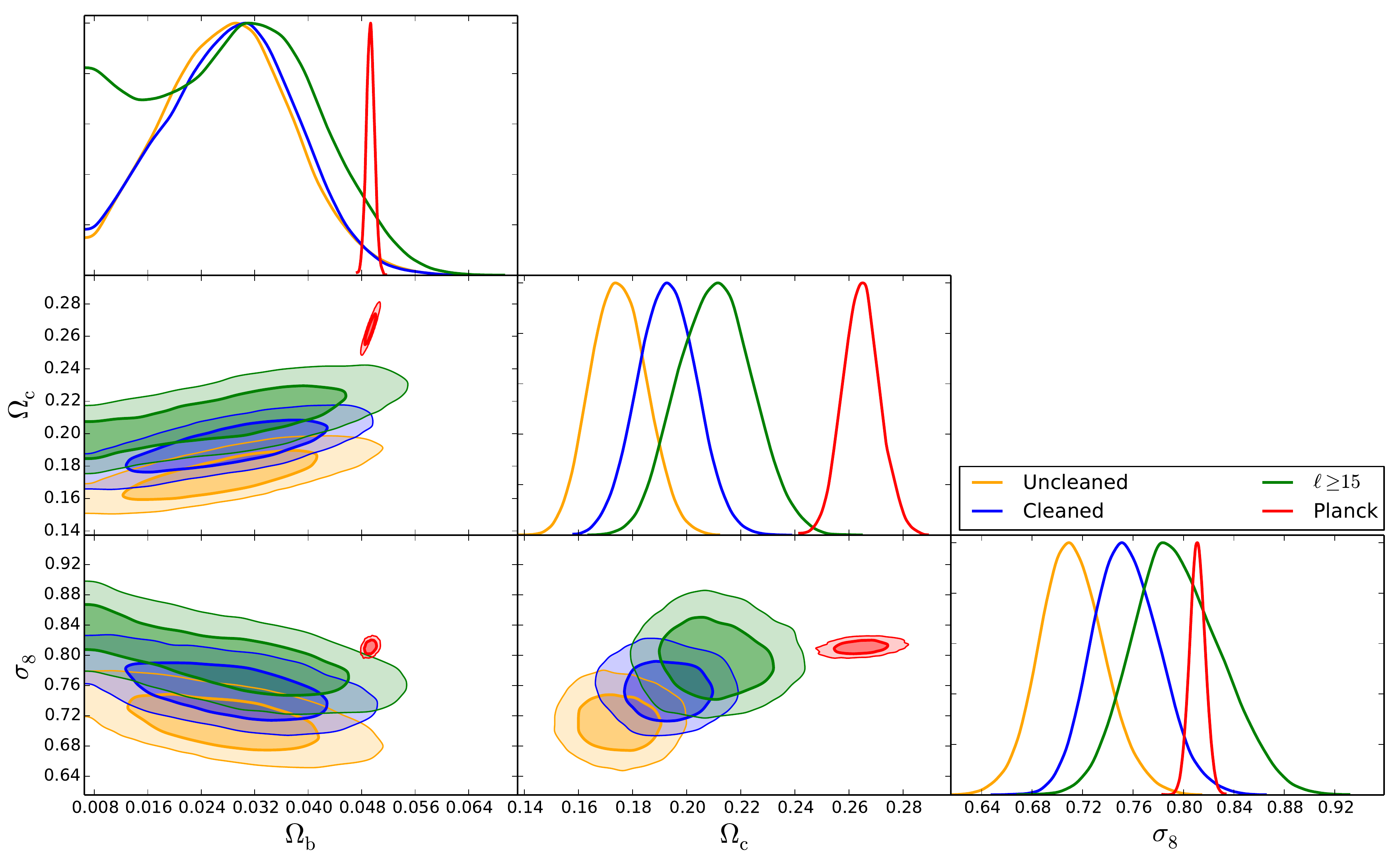}
  \caption{Matrix of posterior plots for the cosmological parameters $\Omega_{\mr{b}}$, $\Omega_{\mr{c}}$ and 
    $\sigma_8$. The diagonal shows the marginal 1D posteriors for each parameter (normalized to their maximum), 
    while the off-diagonal plots show the 2D posteriors, with contours at 68\% and 95\% confidence levels. The orange and blue posteriors 
    represent the results for the $3\leq\ell\leq\ell_{\mr{max}}$ range for the uncleaned and cleaned data, respectively, 
    while the green posteriors show the results for the cleaned data for the $15\leq\ell\leq\ell_{\mr{max}}$ range 
    ($\ell_{\mr{max}}$ is the linear limit given by Table \ref{tab:linear-l}). The plots also show in red the posteriors 
    from Planck baseline.
    }
  \label{fig:data-posterior}
\end{figure}

Figure \ref{fig:data-posterior} shows the 1D and 2D posterior distributions for the cosmological parameters 
that were not directly restricted by Planck priors, for the first three approaches listed in Table \ref{tab:chi2}, and 
compares them with the values according to Planck. In any case, the posteriors disagree with Planck values by 
more than $2\sigma$. The most discrepant parameter is $\Omega_{{\mr{c}}}$, which WSC data push to lower values. 
This is caused by the steep slope of WSC $C_\ell$s, as shown in Fig.~\ref{fig:data-cls}: lower cold matter 
density leads to less small-scale clustering, tilting the power spectrum and making it redder.
Fig.~\ref{fig:data-posterior} also shows that the power reduction at the largest scales produced by 
our cleaning method significantly shifts the posterior closer to the Planck values 
without increasing its size. This demonstrates, using real data, the importance and advantages of our 
cleaning method. The removal of the largest scales ($\ell<15$) moves the posterior even closer to 
the Planck values but increases its width. This suggests that the stellar density 
template may be improved, or that the largest scales are affected by extra effects besides stellar obscuration 
and contamination. In either case, it is clear that identifying and taking these effects into account might be a 
better approach than ignoring the largest scales altogether, as the latter strategy would lead to worse precision.


\section{Robustness tests and search for further systematics}
\label{sec:robustness}

To search for other potential sources of systematic effects in the WSC data, 
we applied a series of modifications to our analysis and checked their impact on the 
measured $C_\ell$s, using the full-sky cleaned $C_\ell$s with $\ell\geq 15$ as 
our starting point. One of these tests, already mentioned in Sec.~\ref{sec:results}, 
consisted in ignoring scales $\ell<21$, with the only impact on the results being that 
of making the posterior distributions wider.

\subsection{Different stellar maps and masks} 
\begin{figure}
  \center
  \includegraphics[width=0.32\textwidth]{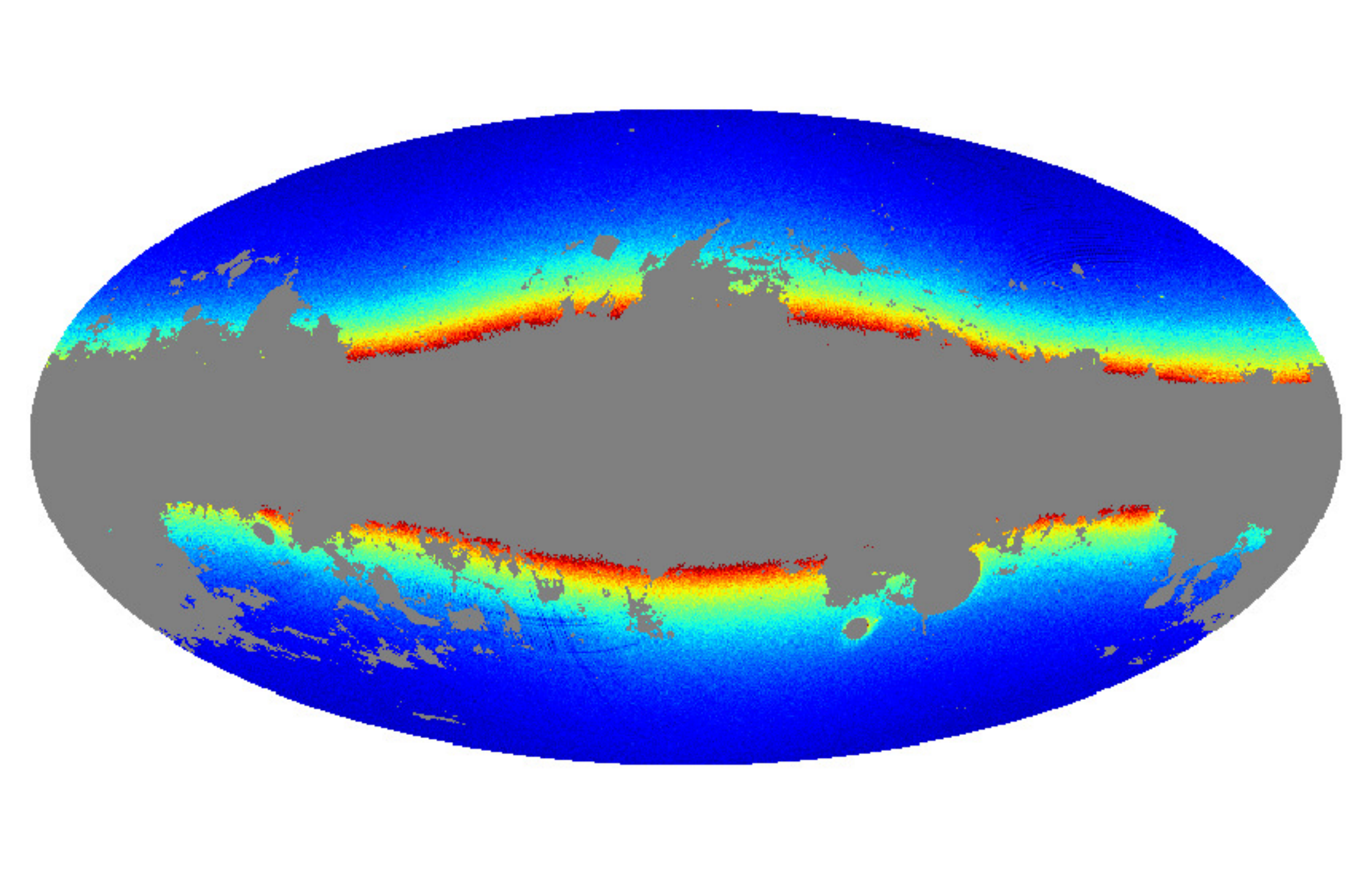}
  \includegraphics[width=0.32\textwidth]{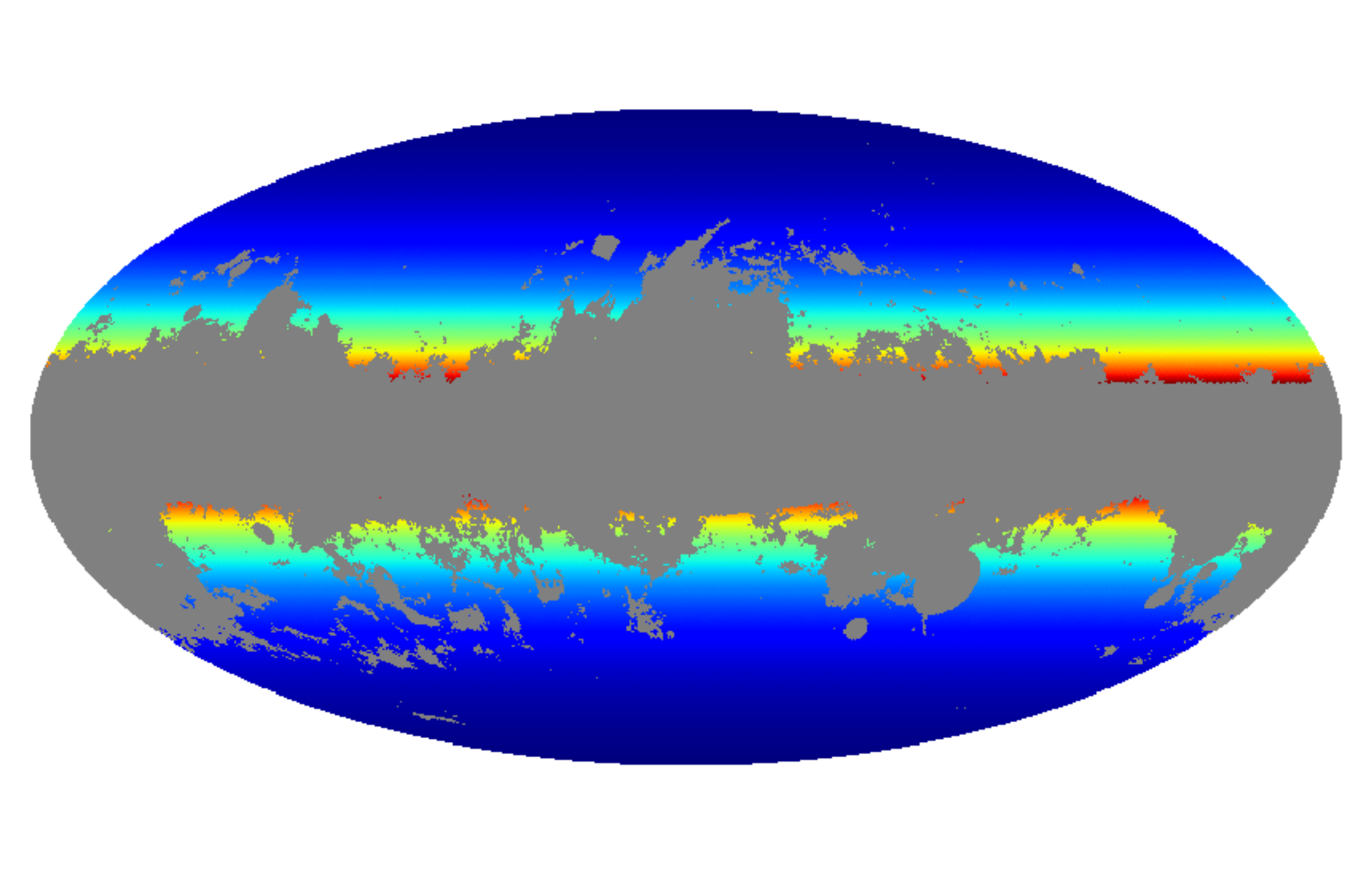}
  \includegraphics[width=0.32\textwidth]{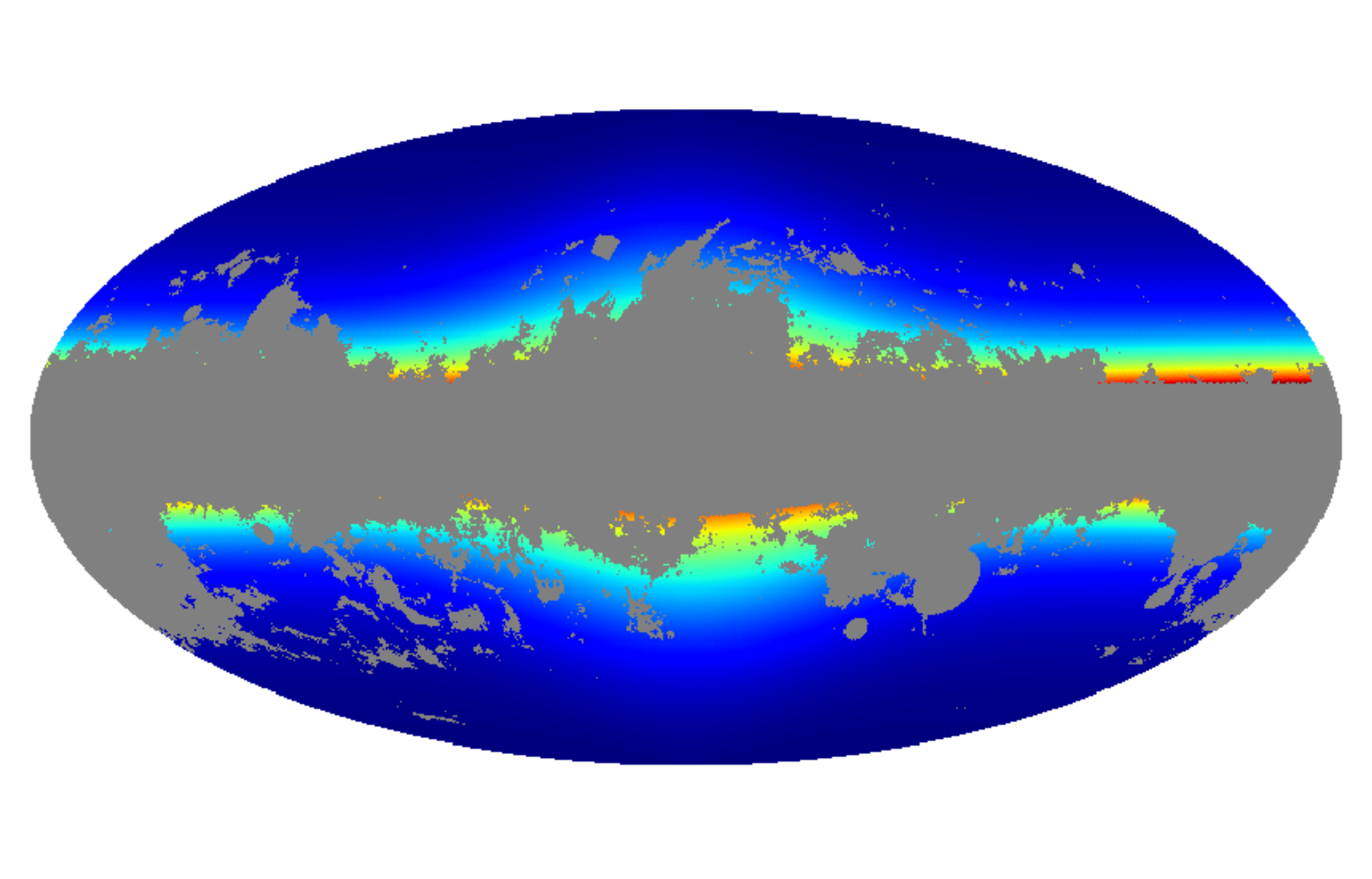}
  \caption{Alternative stellar density templates used to clean the WSC counts maps during the robustness
    tests. From left to right: a Gaia map without smoothing and with a larger mask; a latitudinal exponential 
    model; and a ``disk \& bulge'' model fitted to the cross-match between WSC sources and SDSS stars. 
    The color scales are independent for each map and cover the full template's range.}
  \label{fig:alt-templates}
\end{figure}

When considering possible residual systematics related to stellar density, we first raised two hypotheses, viz. 
(\emph{i}) that problematic regions near the Galactic bulge could be biasing our results, and 
(\emph{ii}) that the Gaussian smoothing applied to the Gaia map could have erased Galactic 
structures at intermediate scales ($\ell\gtrsim30$) that were not removed by our fiducial template. 
In order to verify these claims, we created the template $S(\bm{\theta})$ shown in the left panel of 
Fig.~\ref{fig:alt-templates}, consisting in an unsmoothed map with regions of stellar density three times the mean masked, 
and applied it in the removal of obscuration and contamination effects.

Another possibility we considered was that the map obtained from Gaia was not representing the 
contamination in the WSC. Hence, and based on the cross-match with SDSS presented in Sec.~\ref{sec:contamination}, 
we created a template with the simple exponential model given by Eq.~(\ref{eq:exp-n-stars}) (central panel in 
Fig.~\ref{fig:alt-templates}). Finally, a third template was created, in which the model fitted to the 
matched SDSS stars had an exponential disk (with an extra longitudinal dependence) and a Gaussian bulge 
(Fig.~\ref{fig:alt-templates}, right panel). Note that despite the fact that this approach is based on a direct 
estimate of the contamination, it relies on an extrapolation from the SDSS footprint to the whole sky.

In all cases, the cleaning process reduced the power on the largest scales and only significantly affected 
multipoles $\ell\lesssim20$. The two templates based on Gaia led to very similar outcomes and, for most 
$C^{ij}_\ell$s, improved the results (in terms of large scale power reduction) more than those based on 
the SDSS cross-match.


\subsection{Uncleaned data with $\ell\geq15$}
Although we have shown in Secs. \ref{sec:cl-validation} and \ref{sec:mcmc-validation} 
that our cleaning method does not bias the results (and, in particular, does not tilt the $C^{ij}_\ell$s), 
this demonstration assumed that our stellar density template had the correct shape. To 
test whether our cleaning method could be biasing the results due to a mismatch between the template 
and the true stellar density distribution, we estimated the posterior distribution of the cosmological 
parameters from the uncleaned data, while ignoring the largest scales ($\ell<15$).

Once more, the posteriors remained highly compatible with those for the cleaned $\ell\geq15$ case shown 
in Fig.~\ref{fig:data-posterior}. This testifies that our cleaning method only affects the large-scale 
modes $\ell<15$ and is not responsible for the disagreement between WSC and Planck. 

\subsection{Mixing matrix deconvolution and $H_0$ prior}

As stated in Sec.~\ref{sec:cl}, we estimated the angular power spectra of the unmasked sky by inverting 
Eq.~(\ref{eq:mixing-matrix}), that is, we deconvolved the mixing matrix from the data by multiplying 
the observed pseudo-$C_\ell$s $D_\ell^{ij}$ by the inverse $R_{\ell'\ell}^{-1}$ of the mixing matrix. As this 
process combines data from different scales, it could transfer to smaller scales the power coming from 
systematic effects on the largest scales. An alternative analysis that should not suffer from this effect 
would be to compare the data pseudo-$C_\ell$s (after removing shot-noise and the pixel window function) 
with theoretical $C_\ell$s convolved with the mixing matrix $R_{\ell'\ell}$. If no anomalous power transfer 
happens during the mixing matrix deconvolution in the first approach, both methods should yield the same results.
We ran the simple test of comparing the ratio between deconvolved data $C_\ell$s and fiducial $C_\ell$s with 
the ratio between data pseudo-$C_\ell$s and the same (but convolved) fiducial $C_\ell$s. Both ratios agreed to very good 
extent, demonstrating the robustness of our analysis against forward/backward modeling schemes.

Another analysis choice impact we decided to test was that of the $H_0$ prior. It is known that the Hubble constant 
inferred from Planck ($H_0 = 67.31 \pm 0.96\,\mr{km\,s^{-1}\,Mpc^{-1}}$) is in tension with direct measurements using 
type Ia supernovae ($H_0 = 73.24 \pm 1.74\,\mr{km\,s^{-1}\,Mpc^{-1}}$) \cite{Riess16}. We verified if the Planck 
$H_0$ prior could be responsible for the observed tension between $\Omega_{\mr{c}}$ estimated from WSC and Planck data 
by replacing it with direct measurements prior. Although this change displaced the WSC posteriors, it actually 
increased the tension on $\Omega_{\mr{c}}$ slightly.

\subsection{Redshift distributions}
\label{sec:lorentzian-photo-z}

As mentioned in Sec.~\ref{sec:wsc}, the WSC photo-$z$ errors can also be described by a generalized Lorentzian 
distribution, which results (from Eq.~\ref{eq:spec-z}) in a spec-$z$ distribution in each photo-$z$ bin 
that can also be well modeled by the same distribution. To test whether the Gaussianity 
assumption used in our analysis could bias the results, we employed generalized Lorentzian functions
as spec-$z$ distributions, and used it to generate a model of $C_\ell$s and sample the posterior distribution 
of the constrained parameters. 

Generalized Lorentzian functions have an extra parameter $a$ that controls the importance of its tails 
(see Eq.~\ref{eq:lorentzian}). We ran the MCMC under two configurations: 
(\emph{i}) we let $a_i$ for each $i$ redshift bin vary freely between 1.5 and 4.4; and (\emph{ii}) 
we fixed $a_i$ to the values estimated in \cite{Peacock18}. In both cases, the posterior on the 
cosmological parameters remained quite similar to the one in Fig.~\ref{fig:data-posterior}.

We also investigated the scenario in which ignoring the photo-$z$ correction due to the north-south hemisphere 
asymmetry (used in our main analysis; see \cite{Bilicki16} for details) could bias our results: 
using the photo-$z$ computed directly by ANN$z$, 
we performed again the redshift binning, the cleaning and the measurement of the $C_\ell$. As a result, no significant 
change was observed. 

\subsection{Ignoring cross-$C_\ell$s}

Our main analysis extracted cosmological information not only from the power spectra in each redshift bin 
but also from the cross-correlation between different bins. These cross-$C_\ell$s are more sensitive 
to the spec-$z$ distribution of the galaxies in each bin than the auto-$C_\ell$s, since the 
overlap in the true redshift distributions is one of the main sources of signal in their case. 
To test if the cross-$C_\ell$s could be biasing our results, we repeated the procedure to 
estimate cosmological parameters using only the auto-$C_\ell$s. Although we noticed no significant 
displacement in the mean values, the width of the 1D posterior distributions are larger, showing that the cross-$C_\ell$s do contain extra information.

\subsection{Different color and magnitude cuts}

The WSC catalog originally had a variable color cut dependent on the distance from the Galactic center, which we removed 
by applying a more stringent constant cut all over the sky (see Sec.~\ref{sec:wsc}). We computed the 
WSC $C_\ell$s using this original color cut and verified that it results in more power on the largest 
scales than when using our cut $W1-W2>0.2$, without much change on smaller scales. This effect is 
expected since a position-dependent color cut creates a variable galaxy selection function, 
likely making their expected number and bias also position-dependent. In other words, 
the homogeneity in the number counts discussed by \cite{Bilicki16} is achieved by the original WSC color cut 
not by removing stars at low latitudes but by removing both stars and galaxies.

Another potential source of systematics is angular variations of the survey depth, caused by changes 
in observing conditions (in SuperCOSMOS) or total exposure times (number of observations in WISE). 
To test for the presence of such features, we applied 
more conservative magnitude cuts, going from the WSC fiducial limits $W1<17$, $B<21$ and $R<19.5$ to 
$W1<16.5$, $B<20.7$ and $R<19.1$. Apart from a constant factor increase in power in all $C_\ell$s, 
expected as a consequence of the larger bias usually 
associated with brighter galaxies, no significant change was seen in the angular power spectra.

\subsection{Analysis by hemispheres}
\label{sec:hemispheres}

In the full-sky analysis above we have found that, although our corrections for stellar systematics do shift 
the WSC posterior in the direction of Planck (with respect to the no-correction case), there is still 
considerable discrepancy between WSC and Planck parameters. As there is no significant evidence from 
other LSS studies at similar redshifts that this should be physical \cite{GilMarin15mn,Balaguera17,Abbott18,Loureiro18}, 
here we analyze a possible 
observational effect related to the WSC catalog. Namely, the underlying SuperCOSMOS data are based on 
measurements from two telescopes which had different effective passbands \cite{Hambly01,Peacock16mn}. 
The division between the two `hemispheres' is at $\delta_{1950}=2.5^\circ$. As noted in \cite{Bilicki16}, 
despite color-dependent corrections applied in SuperCOSMOS to match north and south \cite{Peacock16mn}, within 
the WSC flux limits (Sec.~\ref{sec:wsc}) there is still residual difference in mean galaxy surface density, 
the north one being up to 4\% larger than in the south. As this might be a sign of a genuine systematic in 
the data, we decided to perform the same analysis as for the full sky also for the two hemispheres. Similar 
tests were done for 2MPZ \cite{Balaguera17}, which also incorporates SuperCOSMOS as one of the input samples, but 
no significant differences between resulting cosmological parameters were found. We note that the second 
input sample included in WSC -- WISE -- does present non-uniformities in the data but not of hemispherical 
nature, but rather related to the satellite's polar orbit and Moon avoidance maneuvers\footnote{See 
\url{http://wise2.ipac.caltech.edu/docs/release/allsky/expsup/sec3_4.html}}. Since these 
non-uniformities are much more involved, we did not investigate their possible impact on our results.

\begin{figure}
  \center
  \includegraphics[width=1.0\textwidth]{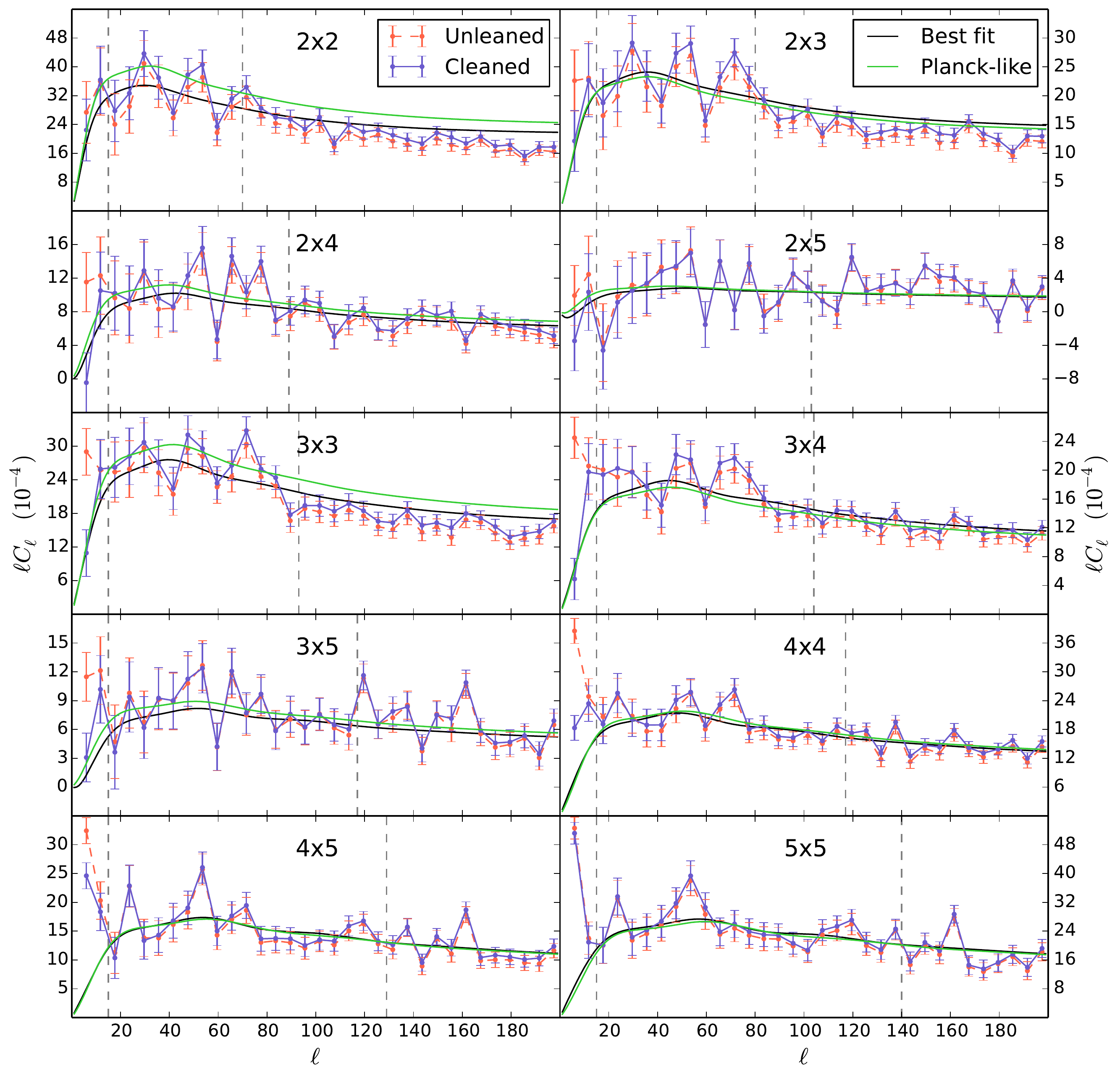}
  \caption{Similar as Fig.~\ref{fig:data-cls}, but for data restricted to the northern Galactic hemisphere. 
      The cleaning process reduces power at $\ell<9$ and re-scales the $C_\ell$s. 
      The best-fit $C_\ell$s to the northern data present slopes that are compatible with Planck.}
  \label{fig:norhtern-cls}
\end{figure}

\begin{figure}
  \center
  \includegraphics[width=1.0\textwidth]{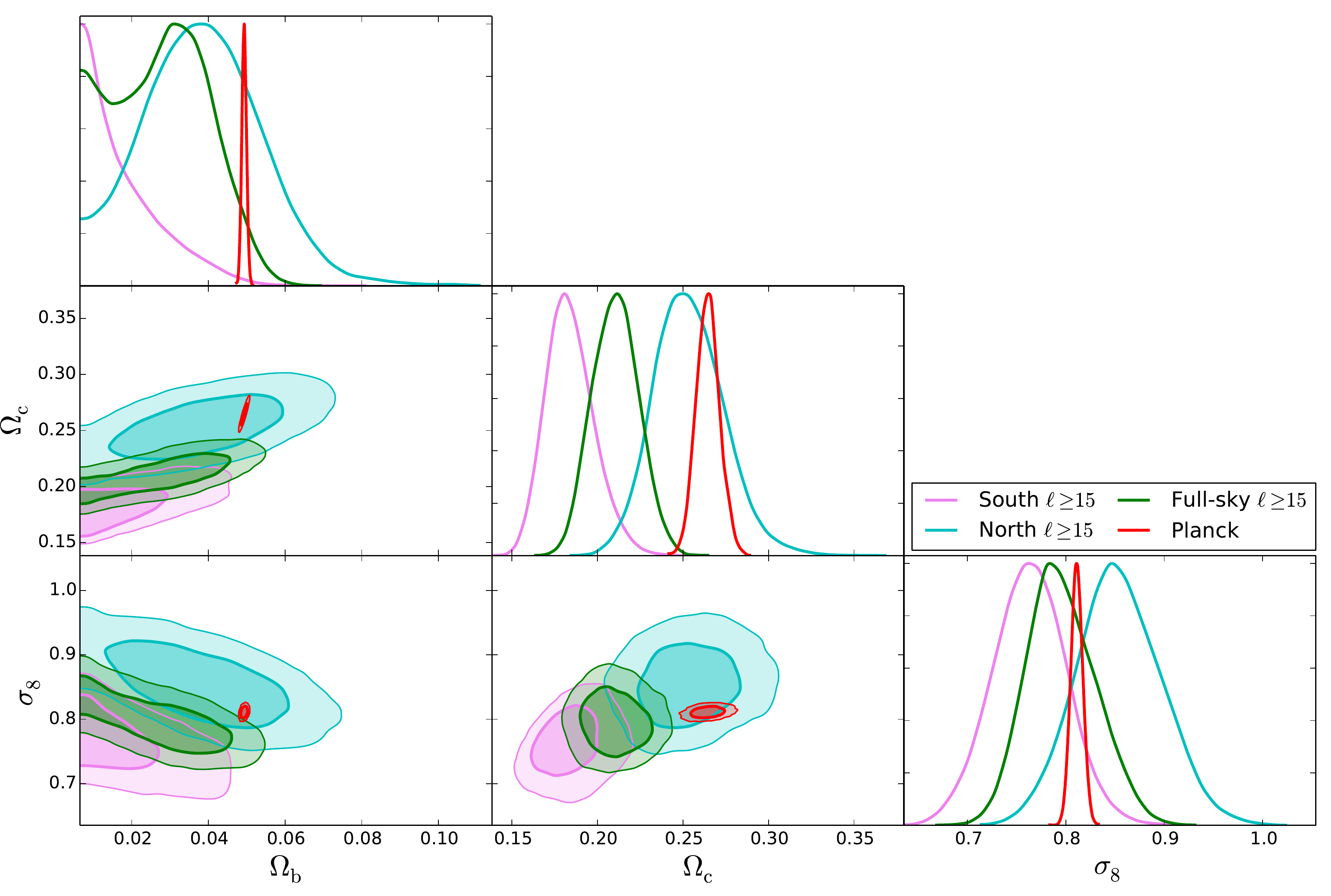}
  \caption{Similar as Fig.~\ref{fig:data-posterior}, but all WSC posteriors employ scales 
    $\ell\geq15$ and use cleaned data. The cyan and pink posteriors were obtained from the northern and southern 
    hemispheres, respectively, while the green posteriors show the results for the full sky.}
  \label{fig:data-posterior-hemisphere}
\end{figure}

Both hemispheres present similar behavior as the full-sky analysis with respect to the 
impact of our cleaning process and removal of multipoles $\ell<15$. When including 
all scales with $\ell\geq 3$ and ignoring contamination, both hemispheres display a 
bad $\chi^2$ and a clear tension with the results from Planck. The cleaning process improves the fit 
and reduces the tension with Planck by lowering the power on the largest scales ($\ell<9$). 
Fig. \ref{fig:norhtern-cls} shows the northern $C_\ell$s before and after cleaning. 
The southern hemisphere -- which 
already presented less excess of power on these scales -- is better improved than the northern 
hemisphere and actually reaches an acceptable $\chi^2$, although both remain inconsistent with Planck. 
When we remove the scales with $\ell<15$, the $\chi^2$ becomes acceptable for both hemispheres, but 
only the northern hemisphere agrees with Planck. This may indicate that different systematics dominate 
each hemisphere. The $\chi^2$ values are shown in Table \ref{tab:chi2} 
and the posterior distributions for each hemisphere under the $\ell\geq15$ cut are shown and compared 
with the full-sky posteriors in Fig.~\ref{fig:data-posterior-hemisphere}. The values of all 13 parameters 
associated with these posteriors are presented in Table \ref{tab:bestfit}.

\begin{table}
\begin{center}
\begin{tabular}{l|cc|cc|cc} 
\hline 
\multicolumn{1}{c|}{} & \multicolumn{2}{|c|}{Full-sky $\ell\geq15$} & \multicolumn{2}{|c|}{North $\ell\geq15$} & \multicolumn{2}{|c}{South $\ell\geq15$} \\
Parameter & Best-fit & Mean $\pm\sigma$ & Best-fit & Mean $\pm\sigma$ & Best-fit & Mean $\pm\sigma$ \\
\hline
$h$ & $0.6703$ & $0.6729(99)$ & $0.6758$ & $0.6729(99)$ & $0.6758$ & $0.6723(97)$ \\
$\Omega_{\mr{b}}$ & $0.0321$ & $<0.0477$ & $0.0340$ & $<0.0648$ & $0.0076$ & $<0.0397$ \\
$\Omega_{\mr{c}}$ & $0.214$ & $0.211(14)$ & $0.251$ & $0.254(20)$ & $0.178$ & $0.184(14)$ \\
$\ln(10^{10}A_{\mr{s}})$ & $3.103$ & $3.088(36)$ & $3.092$ & $3.086(36)$ & $3.078$ & $3.088(36)$ \\
$n_{\mr{s}}$ & $0.9678$ & $0.9652(64)$ & $0.9687$ & $0.9657(63)$ & $0.9638$ & $0.9653(65)$ \\
$b_{2}$ & $1.432$ & $1.424(75)$ & $1.274$ & $1.322(80)$ & $1.394$ & $1.455(82)$ \\ 
$b_{3}$ & $1.284$ & $1.274(63)$ & $1.195$ & $1.229(68)$ & $1.251$ & $1.314(67)$ \\
$b_{4}$ & $1.149$ & $1.144(55)$ & $1.107$ & $1.134(61)$ & $1.104$ & $1.152(58)$ \\
$b_{5}$ & $1.264$ & $1.266(64)$ & $1.254$ & $1.274(75)$ & $1.210$ & $1.238(70)$ \\
$\sigma_{\mr{z},2}$ & $0.0575$ & $0.0581(22)$ & $0.0587$ & $0.0598(36)$ & $0.05591$ & $0.0557(29)$ \\
$\sigma_{\mr{z},3}$ & $0.04970$ & $0.04969(90)$ & $0.0495$ & $0.0498(15)$ & $0.0496$ & $0.0500(12)$ \\
$\sigma_{\mr{z},4}$ & $0.04401$ & $0.04400(73)$ & $0.0449$ & $0.0453(11)$ & $0.0436$ & $0.0439(10)$ \\
$\sigma_{\mr{z},5}$ & $0.0353$ & $0.0354(13)$ & $0.0384$ & $0.0379(18)$ & $0.0344$ & $0.0333(17)$ \\
$\sigma_8$ & $0.7915$ & $0.796(35)$ & $0.869$ & $0.854(43)$ & $0.792$ & $0.763(38)$ \\
\hline 
\end{tabular} 
\end{center}
\caption{Best-fit values of the parameters of a $\Lambda$CDM model and the mean and 
    standard deviation of their marginal posteriors for the full-sky, northern and southern datasets. All 
    datasets shown here were cleaned and fit from $\ell\geq15$ to linear scales. Since the posterior of $\Omega_{\mr{b}}$ 
    reaches the  \texttt{CLASS} limit $\Omega_{\mr{b}}=0.0065$, we quote its 95\% confidence upper limit instead of 
    the mean.}
\label{tab:bestfit}
\end{table}

\section{Conclusions and summary}
\label{sec:conclusions}

In this work we have tested a method for estimating and correcting for stellar obscuration and contamination 
in galaxy redshift catalogs, aiming at extracting cosmological information therefrom. In particular, we have 
implemented it on the WISE$\times$SuperCOSMOS (WSC) photo-$z$ dataset \cite{Bilicki16}. 
We first exposed in Sec.~\ref{sec:character} the presence of both systematic effects in WSC, a 
feature that will be common -- albeit with a smaller amplitude -- in every photometric survey, including LSST, 
J-PAS and Euclid. In that Section we also proposed a model for this feature that 
ties both systematic effects together with a single stellar density template (Eq.~\ref{eq:sources-model}), 
and showed that such an assumption is supported by the data (Figs. \ref{fig:wsc-dev-b} and \ref{fig:wsc-dev-gaia}). 

The fact that both effects -- obscuration and contamination -- are linked by the template demands 
a strategy other than simple correlation \cite{Rybicki92,Elsner17} to quantify them. We proposed 
the usage of variance measurements in bands of constant stellar density as a way of determining obscuration 
(Eq.~\ref{eq:iso-model}), which later can be used to estimate the contamination and real galaxy density
(Eqs. \ref{eq:compute-beta} and \ref{eq:compute-ng}). We verified that this strategy is able to detect and 
quantify all these parameters with less than 10\% bias (Table \ref{tab:recov-pars}). Although Figs. 
\ref{fig:sim-cls} and \ref{fig:sim-mcmc-marginal} show that the effect of any residual bias is negligible 
in the rest of our analysis, we propose that this cleaning method can be improved if we take into 
account that galaxy densities are spatially correlated.

  It is worth pointing out that one may disentangle the effects of obscuration and contamination directly on the 
  power spectra -- since they result in different signatures (see Sec. \ref{sec:cl-validation}) -- 
  by treating them as nuisance parameters and fitting them together with the cosmology.
  In this process, it is relevant to not only analyse the data $C_\ell$s but also the cross power
    spectra between data and stellar density template. The contamination level may be better constrained by it,
    whereas the measurement of obscuration might benefit from a cross-$C_\ell$ between the template and the
    observed density difference squared, $\delta n_{\mr{obs}}^2$. Testing this approach should be part of a future work.

All of our methodology was validated by applying it to simulations created by the FLASK code \cite{Xavier16} 
and comparing the measured power spectrum and the constrained parameters with their true input values 
(see Sec.~\ref{sec:validation}). That analysis demonstrated that applying our 
cleaning method to data containing obscuration and contamination significantly improved the cosmological 
parameter inference. This method was also applied to real WSC data, and Table \ref{tab:chi2} and 
Fig.~\ref{fig:data-posterior} show that it improved the power spectra fit at $\ell<15$ and increased the agreement 
between the cosmology derived from WSC power spectra and from the Planck satellite. 

Despite the improvement obtained both in the full-sky and in separate hemispheres by removing 
contamination and obscuration, the constraints derived from the angular clustering of WSC still disagree with 
those derived from the Planck mission if we include the largest scales. In fact, the minimum $\chi^2$ shows that the 
standard model is a bad fit to the data in the full-sky and northern hemisphere case (see Table \ref{tab:chi2}) 
even for parameter values significantly different from Planck. The fit is significantly improved 
if we ignore multipoles $\ell<15$ (reaching a $p$-value of $\sim$6\% for the full-sky case, which could be considered 
adequate), proving that much of the deviation between data and model happens at the largest scales. 
The reasons for this excess of power is still unclear and should be further investigated. Possible explanations 
are that (\emph{i}) stellar density templates, used to remove contamination, must be improved to 
better represent WSC true distribution; and that (\emph{ii}) important systematic effects not related 
to stars -- possibly Galactic extinction and photometric/observational errors -- need to be taken into account. 
We do point out, however, that we have tested three other templates (Fig.~ref{fig:alt-templates}) and they 
have led to similar (or more biased) results. 

As the scales $\ell\sim15$ are close to the power spectrum turn-over at redshifts probed by WSC, we could claim  
that the excess of power at scales larger than this comes from physical effects not considered by our 
cosmological model, e.g.: by non-Gaussian initial conditions caused by certain inflationary models 
\cite{Dalal08}; by the fact that galaxy forms at matter density peaks, which could lead to a scale-dependent 
bias \cite{Durrer03}; or by alternative dark energy models that produce different redshift space distortions 
\cite{Takada06b}. These effects, however, are expected to be small and probably should wait for the investigation 
of other possible systematics sources. It is worth pointing out that an excess of power is constantly seen on the 
largest scales of galaxy surveys \cite{Thomas11,Balaguera17,Loureiro18}, and due to its correlation with 
stellar density and other reasons, these modes are mostly discarded. This is unfortunate as the physical effects 
mentioned above remain untested.

We have tested our data for several potential sources of 
systematics using the full-sky data limited to $\ell\geq 15$ as a reference. In Sec.~\ref{sec:robustness} 
we verified that our results are not affected by photo-$z$ error distributions, cleaning 
strategy, color and magnitude cuts, photo-$z$ estimations, mixing matrix deconvolution, 
the use of cross-$C_\ell$s and the restriction to $\ell\geq 21$. We did find, however, a hemispherical pattern in the data.
Fig.~\ref{fig:data-posterior-hemisphere} shows that the independent analyses of the two equatorial hemispheres
lead to incompatible cosmological parameters: while the analysis restricted to $\ell\geq 15$ and to the 
northern hemisphere agrees with Planck (with $\Omega_{c}=0.254^{+0.018}_{-0.022}$, $\sigma_8=0.854^{+0.042}_{-0.045}$ and 
$\Omega_{\mr{b}}<0.065$ at 95\% confidence limit 
when combined with Planck priors on $H_0$, $A_s$ and $n_s$), the same analysis in the southern hemisphere does not. 
This difference in extracted cosmological parameters is traced back to different slopes in the measured power spectra, 
with the southern and full-sky $C_\ell$s being steeper than the northern one.
Considering it is unlikely that such a large difference has a cosmological origin -- especially since cosmological parameters 
have been extracted from the southern hemisphere before and shown to agree with Planck 
\cite{Balaguera17,Abbott18} -- this indicates that systematics affecting scales $\ell\gtrsim 15$ 
are located in the southern part of WSC data.

Investigating what is the actual source of these systematics is beyond the scope of the present work. 
One possible explanation is residual miscalibration of SuperCOSMOS data in the south. Namely, 
the input photographic material for that sample was based on observations from two telescopes (POSS-II 
in the north and UKST in the south) with 
effectively different passbands. Despite the efforts to match the passbands using specific color terms, 
anchored to SDSS as the calibrator \cite{Peacock16mn}, differences may still exist. As the WSC sample 
is selected using maximum apparent magnitude cuts, in particular in the SuperCOSMOS $B$ and $R$ bands, 
this may propagate into inconsistent sample selections in the two hemispheres. Indeed, already in 
\cite{Bilicki16}, a difference in effective source densities between north and south was reported. 

However, the most plausible explanation for the north-south difference in large-scale power comes from 
residual uncertainties in photometric zero points. Where direct SDSS calibration was lacking, 
plate zero points were given an initial estimate from mean optical-2MASS color, and these values 
were then perturbed in an attempt to optimize agreement in plate overlap zones. It seems likely 
that this process will yield zero points that are imperfect at the level of a few hundredths of a 
magnitude \cite{Skrutskie06mn}, imprinting a pattern of $5\times5\,\mr{deg^2}$ tiles on the galaxy 
distribution with an rms variation in surface density of several percent. This is the right order of 
magnitude to cause the offsets that we measure (J. Peacock, private communication).
The second input dataset of WSC -- WISE -- is unlikely to be at the origin of the 
hemispherical differences as any non-uniformity in that survey is related mostly to telescope's polar 
orbit and Moon avoidance strategy. Thus, our present findings seem to suggest that the observed asymmetry 
is due to residual systematics in the UKST data.

\section*{Acknowledgments}

The authors thank Vinicius Placco and Laura Sampedro for advice on the use of Gaia data, Filipe Abdalla for comments 
related to power spectrum measurements and cosmological parameter extraction and John Peacock for valuable information on 
SuperCOSMOS and useful advice on data analysis and interpretation. 

We thank the Wide Field Astronomy Unit (WFAU) at the Institute for Astronomy, Edinburgh, for storing the 
WISE$\times$SuperCOSMOS data, available for download from \url{http://ssa.roe.ac.uk/WISExSCOS}.
We have also made use of data from the European Space Agency (ESA)
mission {\it Gaia} (\url{https://www.cosmos.esa.int/gaia}), processed by
the {\it Gaia} Data Processing and Analysis Consortium (DPAC,
\url{https://www.cosmos.esa.int/web/gaia/dpac/consortium}). Funding
for the DPAC has been provided by national institutions, in particular
the institutions participating in the {\it Gaia} Multilateral Agreement.
Funding for the Sloan Digital Sky Survey IV has been provided by the Alfred P. Sloan Foundation, 
the U.S. Department of Energy Office of Science, and the Participating Institutions. SDSS-IV acknowledges
support and resources from the Center for High-Performance Computing at
the University of Utah. The SDSS web site is \url{http://www.sdss.org}.
SDSS-IV is managed by the Astrophysical Research Consortium for the 
Participating Institutions of the SDSS Collaboration including the 
Brazilian Participation Group, the Carnegie Institution for Science, 
Carnegie Mellon University, the Chilean Participation Group, the French 
Participation Group, Harvard-Smithsonian Center for Astrophysics, 
Instituto de Astrof\'isica de Canarias, The Johns Hopkins University, Kavli Institute 
for the Physics and Mathematics of the Universe (IPMU) / 
University of Tokyo, the Korean Participation Group, Lawrence Berkeley National Laboratory, 
Leibniz Institut f\"ur Astrophysik Potsdam (AIP),  
Max-Planck-Institut f\"ur Astronomie (MPIA Heidelberg), 
Max-Planck-Institut f\"ur Astrophysik (MPA Garching), 
Max-Planck-Institut f\"ur Extraterrestrische Physik (MPE), 
National Astronomical Observatories of China, New Mexico State University, 
New York University, University of Notre Dame, 
Observat\'ario Nacional / MCTI, The Ohio State University, 
Pennsylvania State University, Shanghai Astronomical Observatory, 
United Kingdom Participation Group,
Universidad Nacional Aut\'onoma de M\'exico, University of Arizona, 
University of Colorado Boulder, University of Oxford, University of Portsmouth, 
University of Utah, University of Virginia, University of Washington, University of Wisconsin, 
Vanderbilt University, and Yale University.

This work has made use of the computing facilities of the Laboratory of Astroinformatics 
(IAG/USP, NAT/Unicsul), whose purchase was made possible by the Brazilian agency FAPESP (grant 2009/54006-
4) and the INCT-A. 
HSX acknowledges 
FAPESP Brazilian funding agency for the financial support.
ABA acknowledges financial support from the Spanish Ministry of Economy and Competitiveness (MINECO) 
under the Severo Ochoa program SEV-2015-0548.
MB is supported by the Netherlands Organization for Scientific Research, NWO, through grant number 614.001.451 and by the Polish Ministry of Science and Higher Education through grant DIR/WK/2018/12. 

\bibliographystyle{JHEP}
\bibliography{main}

\end{document}